\documentclass[final,11pt]{elsarticle}

\RequirePackage{fix-cm}
\usepackage{bm}
\usepackage[misc]{ifsym}
\usepackage{graphicx}
\usepackage{mathptmx}      

\usepackage{enumerate}
\usepackage[notcite,notref]{showkeys}
\usepackage{lipsum}
\usepackage{widetext}
\usepackage{url}

\usepackage{multicol}        
\usepackage[bottom]{footmisc}

\usepackage{caption}
\usepackage{algorithm}
\usepackage{algorithmic}
\usepackage{subfigure}
\usepackage{amsmath,amssymb}
\usepackage{graphicx}        
\usepackage{cases}
\usepackage{tikz}
\usetikzlibrary{positioning,decorations,arrows,patterns,shapes,backgrounds}

\usepackage{array}

\makeindex             

\newcommand{\x}{\mathbf{x}}

\newcommand{\w}{b}
\newcommand{\nw}{{c}}
\newcommand{\cw}{w}
\newcommand{\cn}{{n}}

\newcommand{\bV}{{\bf V}}
\newcommand{\bJ}{{\bf J}}
\newcommand{\bD}{{\mathcal{D}}}

\newcommand{\mB}{\mathcal{B}}

\newcommand{\tL}{\text{L}}

\newcommand{\CO}{\text{CO}_2}
\newcommand{\mol}{\text{mol}}

\newcommand{\salt}{\text{salt}}

\newcommand{\absK}{{\bf K}}
\newcommand{\g}{\text{\bf g}}
\newcommand{\degree}{{\ensuremath{^\circ}}}
\newcommand{\di}{\mathrm{div}}
\newcommand{\dt}{\partial_{t}}
\newcommand{\dd}{\mathrm{d}}

\newcommand{\mF}{\mathcal{F}}
\newcommand{\mN}{\mathcal{N}}
\newcommand{\mU}{\mathcal{U}}

\newcommand{\pro}{\mu}
\renewcommand{\dim}{d}
\newcommand{\om}{\omega}

\renewcommand{\to}{\rightarrow}

\newcommand{\R}{\mathbb{R}}

\renewcommand{\vec}[1]{{\bm{#1}}}
\newcommand{\var}[1]{{\mathrm{var}}\left\{ {#1} \right\}}

\newcommand{\avemu}[1]{\mathrm{E}\left\{{#1}\right\}}

\renewcommand{\xi}{\theta}
\newcommand{\xxi}{\vec{\xi}}
\newcommand{\ip}[2]{\left( {#1}, {#2} \right)}

\newcommand{\norm}[1]{\left\| {#1} \right\|_{\scriptscriptstyle L^{2}(\Omega)}}

\newcommand{\myvert}{\vert}
\newcommand{\xxii}{\xxi^{\scriptscriptstyle{[-i]}}}
\newcommand{\Qleak}{Q_\text{leak}}
\newcommand{\Tarrival}{t_\text{arrival}}
\newcommand{\Qmax}{Q_\text{leak}^\text{max}}
\newcommand{\Tmaxleak}{t_\text{maxleak}}
\newcommand{\defeq}{\mathrel{\mathop:}=}

\RequirePackage[normalem]{ulem} %
\RequirePackage{color}\definecolor{RED}{rgb}{1,0,0}\definecolor{BLUE}{rgb}{0,0,1} %
\begin{document}

\begin{frontmatter}

\title{Probabilistic modeling and global sensitivity analysis for $\CO$ storage in geological formations: a spectral approach}


\author[kaust,ices]{Bilal~M.~Saad}
\ead{bilal.saad@kaust.edu.sa}
%
\author[ncsu]{Alen Alexanderian\corref{cor1}}
\ead{alexanderian@ncsu.edu}
\author[mont]{Serge~Prudhomme}
\ead{serge.prudhomme@polymtl.ca}
\author[kaust,duke]{Omar~M.~Knio}
\ead{omar.knio@duke.edu}

\cortext[cor1]{Corresponding author}

\address[kaust]{King Abdullah University of Science and Technology, Division of Computer, Electrical and Mathematical Sciences \& Engineering, 
4700 KAUST, Thuwal 23955--6900, Kingdom of Saudi Arabia}
\address[ices]{The Institute for Computational Engineering and Sciences, The University of Texas at Austin, Austin, TX, USA}
\address[ncsu]{Department of Mathematics, North Carolina State University, Raleigh, NC, USA}
\address[mont]{Department of Mathematical and Industrial Engineering, \'Ecole
Polytechnique de Montr\'eal, Montr\'eal, Canada}
\address[duke]{Department of Mechanical Engineering and Materials Science,
Duke University, Durham, NC 27708, USA}

\begin{abstract}
This work focuses on the simulation of $\CO$ storage in deep underground
formations under uncertainty and seeks to understand the impact of
uncertainties in reservoir properties on $\CO$ leakage.
To simulate the process, a non-isothermal two-phase two-component flow system
with equilibrium phase exchange is used.  Since model evaluations are
computationally intensive, instead of traditional Monte Carlo methods, we rely
on polynomial chaos (PC) expansions for representation of the stochastic model
response.  A non-intrusive approach is used to determine the PC coefficients.
We establish the accuracy of the PC representations within a reasonable error
threshold through systematic convergence studies.  In addition to
characterizing the distributions of model observables, we compute probabilities
of excess $\CO$ leakage.  Moreover, we consider the injection rate as a
design parameter and compute an optimum injection rate that ensures that the
risk of excess pressure buildup at the leaky well remains below acceptable
levels. 
We also provide a comprehensive analysis of sensitivities of $\CO$ leakage,
where we compute the contributions of the random parameters, and their
interactions, to the variance by computing first, second, and total order Sobol' indices. 
\end{abstract}


\end{frontmatter}

%
%
%

\date{Received: date / Accepted: date}

%

%
%
\section{Introduction}\label{sec:introduction}
Carbon capture and storage (CCS) is an important topic related to the reduction of
$\CO$ pollution in the atmosphere.  In general, CCS is process of capture and
long-term storage of $\CO$. Different variants for $\CO$ storage are being explored,
with the storage in deep underground formation such as oil fields, gas fields,
abandoned mines, and saline formations being of highest interest. Various risks
exist in $\CO$ sequestration  in deep underground formations, the most important
being (i) $\CO$ leakage through caprock failure, faults, and abandoned wells; (ii)
structural failure due to large pressure peaks; and (iii) brine displacement and
infiltration into drinking water aquifers. Quantification of the risks is of
paramount importance for decision makers when evaluating the storage
approaches before this technology can be implemented on large scale projects.
 In the case of deep geological storage of
$\CO$, there have been significant research efforts dealing with
mathematical and numerical models for simulating the $\CO$ injection processes
into geological formations. Nordbotten et al.
\cite{Nordbotten2004,Bench11a,Bench11b}, presented in a series of papers the
development of a semi-analytical model to describe the space and time evolution
of $\CO$ plumes and the leakage through abandoned wells.  A reduced spatial
dimension model based on vertical equilibrium was discussed by Nilsen et al.
\cite{Halvor2011}.  Ebigbo et al. \cite{Ebigbo07} set up 
benchmark examples in order to compare different modeling approaches such as
numerical and semi-analytical models, for the problem of $\CO$ leakage.
Class et al.~\cite{Bench1} published a benchmark study, comparing a number of
mathematical and numerical models with different complexities for 
problems related to $\CO$ storage in geologic formations.

$\CO$ sequestration is a complex multiphysics process, in which multiphase
multicomponent flows play a critical role. The fact that the $\CO$ should be
stored for many thousands of years implies that full scale experiments are not
possible, and computer simulation is the main approach for exploring the
feasibility of different $\CO$ storage options. 
However the mathematical models of underground $\CO$ storage involve many
sources of geological uncertainties \cite{Hansson2009,Class2009}.  These
uncertainties are due to the limited knowledge about reservoir properties such
as porosity and permeability.
These sources of uncertainty lead to large variabilities in the
predictive modeling of subsurface processes. Hence, one needs to propagate such
uncertainties throughout the calculations to quantify their impact on results
of computer simulations. This requires the use of stochastic modeling
approaches.

\paragraph{Survey of literature on uncertainty quantification (UQ) for $\CO$ storage}
In~\cite{Sun2013} the authors utilize a stochastic response surface method for
assessment of leakage detectability for $\CO$ sequestration, by parameterizing
the spatially heterogeneous reservoir permeability using Karhunen--Lo\`eve
expansion. However, they used the analytical solution developed by Nordbotten
et al. \cite{Bench11a} to generate the pressure distribution at the injection
zone, which is then used to calculate the leakage flux into a confined aquifer
using Darcy's law.  The analytical solution in~\cite{Bench11a} assumes that the
phase saturations and fluid viscosities are constant within each zone, that the
capillary effects are small, and that vertical equilibrium applies to the
entire flow system.  In~\cite{Oladyshkin2011}, the authors use polynomial chaos (PC) expansions
for probabilistic analysis of the $\CO$ leakage rate in the $\CO$ benchmark
presented by Class et al.~\cite{Bench1}.  In that article, the authors use a
number of simplifying assumptions to set up the mathematical model: fluid
properties such as density and viscosity are constant, all processes are
isothermal, $\CO$ and brine are immiscible phases, capillary pressure is
negligible and mutual dissolution is neglected.

The article \cite{walter2012} provides estimates of the risk of brine discharge
into freshwater aquifers following $\CO$ injection into geological formations
and resultant salt concentrations in the overlying drinking water aquifers
using arbitrary PC expansions combined with the probabilistic collocation method of
~\cite{Oladyshkin2011b}.  Other works
include~\cite{Oldenburg2008,Oldenburg2009} where the authors develop a
screening and ranking method and a certification framework based on effective
trapping for geologic carbon sequestration, for selecting suitable storage
sites on the basis of health, safety, and environmental (HSE) risk resulting
from $\CO$ or brine leakage.  Similarities and differences between radioactive
waste disposal and $\CO$ storage for performance assessment have been discussed
in~\cite{Maul2007}. We also mention the paper~\cite{Quanlin2008} that presents a simple
analytical method for the quick assessment of the $\CO$ storage capacity in
closed and semi-closed systems to assess the expected pressure buildup and
$\CO$ storage capacity in such potentially pressure-constrained systems.
 
\paragraph{Spectral methods for UQ} In the present
work, we rely on spectral UQ methods to build a surrogate model for the
nonlinear function that maps the uncertain model parameters to the model
observables. In particular, we utilize PC expansions to
build such surrogates. PC expansions, whose theory goes back to the late 30's and
40's~\cite{Wiener:1938,Cameron:1947},   have become an increasingly popular
tool in recent years as they provide efficient means for performing UQ in
computationally intensive mathematical models; see e.g.,
\cite{Xiu:2002,Xiu:2002b,Xiu:2003,LeMaitre:2004a,LeMaitre:2004b,maitre:2010,Oladyshkin2011,alen:2012,alexanderian2012global,WinokurConradSrajEtAl13,YanGuoXiu12,PengHamptonDoostan,Bigoni2016}
for a nonexhaustive sample of research contributions to numerical methods for UQ
using PC expansions and applications of these methods to real world problems.

PC methods employ an approximation of the model variables in terms of
a spectral expansion in an orthogonal polynomial basis. Once available, the PC
representations can be used to efficiently approximate the statistical
properties of the model outputs.  Generally, there are two approaches for
computing a PC expansion: (1) intrusive methods (see e.g.~\cite{Xiu:2002,Xiu:2002b,Xiu:2003,LeMaitre:2004a,LeMaitre:2004b,maitre:2010}) 
and (2) non-intrusive
methods (see e.g.,~\cite{Babuska:2007,maitre:2010,alen:2012}). 
Intrusive methods require a reformulation of the original
uncertain partial differential equations (PDEs) that govern the system, through a Galerkin projection onto the
PC basis \cite{Ghanem:1991a,knio:2010}.  This entails the need for rewriting
the existing  deterministic solvers.  Subsequently, one has to solve a larger
system for the time/space evolution of the PC coefficients. Non-intrusive
methods, on the other hand, provide a means to compute the spectral
representation via a sampling of the  existing deterministic solvers.  In this
paper, we will follow a non-intrusive approach to compute the coefficients in the
PC expansion. 

\paragraph{Our approach and contributions}
Existing analyses of uncertainties in CCS using PC expansions, either rely on
simplified physical models or do not rigorously establish the accuracy of the
PC representations of the model observables used for uncertainty analysis.  The
goal of this article is to further the understanding of the impact of
parametric uncertainties in the physical processes involved in CCS by using a
more comprehensive physical model, a rigorous numerical study of the accuracy
of the computed PC representations for the quantities of interest, and a
comprehensive analysis of the impact of parametric uncertainties in the
physical processes involved in $\CO$ storage, in the benchmark geological
structure under study.

The fluid properties such as density, viscosity, and enthalpy of the $\CO$ and
brine phases are expected to change as the $\CO$ rises, affecting strongly
$\CO$ arrival time to the leaky well and the leakage rate value of the $\CO$.
Therefore, we model these fluid properties as functions of the aquifer
conditions, and use a non-isothermal two-phase two-component model to describe
the flow processes of the leakage problem.  In addition, we use nonlinear
functions for the capillary pressure and the relative permeability for each
phase.  In Section~\ref{sec:PDEs}, we outline the benchmark problem, and
describe in detail the governing PDEs, our modeling assumptions, as well as the
numerical solver used.  

We rely on PC representations (see section~\ref{sec:spectralUQ} for the
background material) to  propagate the uncertainties in reservoir
absolute permeability, permeability of the leakage well and reservoir porosity,
and in the injection rate on model observables of interest; see
Section~\ref{sec:UQsetup} for the description of the statistical model for
uncertain parameters, and definition of the quantities of interest.  

In section~\ref{sec:numerics}, we present a comprehensive analysis of the
impact of parametric uncertainties in the physical processes involved in $\CO$
storage, in the benchmark geological structure under study.  A novel feature of
the present work is a statistical analysis of the arrival time of the $\CO$
plume at the leaky well. This is important because monitoring $\CO$ arrival time in leaky wells
and/or in observations wells is a key factor for successful storage
management to reduce risk of leakage and contamination of subsurface resources.
In addition, we study the uncertainties in $\CO$ leakage
through the leaky well, the maximum leakage ratio and the time the maximum is
attained, as well as the caprock pressure. 
A detailed computational study of the convergence of the PC representations
in distribution as well as in the sense of $L^2$ is conducted.
Moreover, using a hierarchy of quadrature rules of different resolutions, we establish that
the quadrature rule used to compute PC representations of the model observables
has sufficient accuracy.  Performing such convergence studies, which is sometimes
omitted in applications of spectral UQ methods in uncertainty quantification
for $\CO$ storage, is a crucial first step that establishes the accuracy and
suitability of the PC surrogate for the analysis that follows. 
We use the computed PC representations to understand the time dependent behavior of the
statistical distribution of selected quantities of interest (QoIs).
We also consider the statistical response of the
caprock pressure to the uncertain parameters, and devise a method for choosing
an optimal injection rate that results in minimal risk of excess pressure
buildup at the leaky well. 

To further understand the impact of the uncertain parameters, we provide a
comprehensive analysis of sensitivity of $\CO$ leakage to the uncertain
parameters (see Section~\ref{sec:sensitivity_analysis}).  
This is achieved by computing the Sobol'
indices~\cite{Sobol:1990,Homma:1996,Sobol:2001,Saltelli:2002}.  Traditional
methods for computing the Sobol' indices rely on computationally expensive
sampling-based methods that require thousands of model evaluations.  On the
other hand, PC expansions provide an efficient means to compute the Sobol' sensitivity
indices~\cite{Crestaux:2009,Sudret:2008,alexanderian2012global}; see 
also~\cite{Sergey2013,Namhata2016,Sergey2017} for application of PC-based
sensitivity analysis to $\CO$ storage.
We find that all the uncertain parameters under study have a significant impact
on model variability, but that the balance of sensitivity indices changes over
time.  We also quantify the impact of the interactions between the uncertain
parameters to variance.  To this end, we compute second-order (joint)
sensitivity indices that quantify the impact of pairwise interactions between
the parameters.  Moreover, to shed further light into the impact of overall
interactions among uncertain parameters to the variance, we introduce a
modified variance-based sensitivity measure, which we call the \emph{mixed} index.
This mixed index, which we describe in detail below, can be computed at
negligible  computational cost, once a PC surrogate is available.  We also
analyze the global sensitivity of the $\CO$ saturation to the uncertain inputs
over the three-dimensional computational domain.

%
%

\section{The mathematical model}\label{sec:PDEs}

\subsection{Description of the benchmark problem}
\begin{figure*}[ht]
  \centering
  \includegraphics[width=1\textwidth]{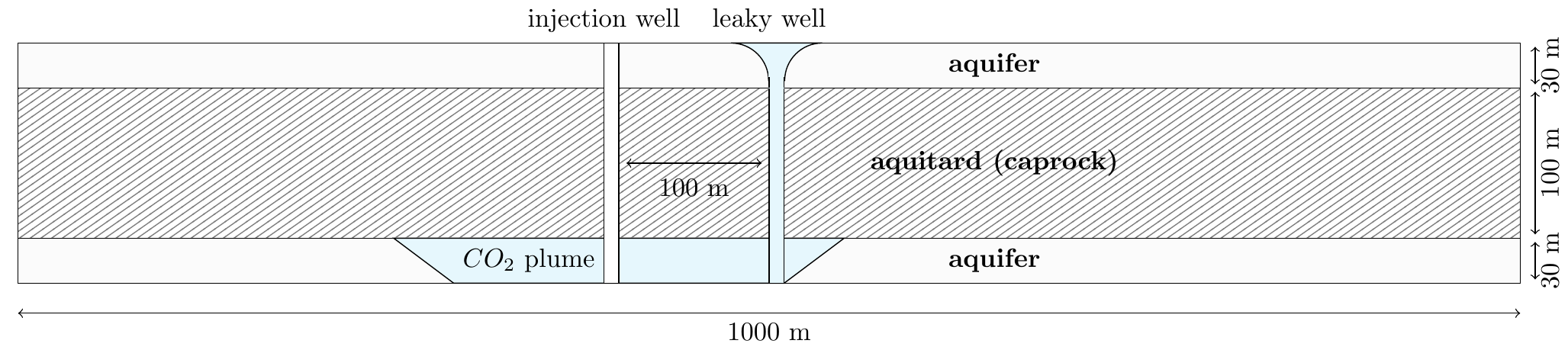}
  \caption{Schematic view of the benchmark problem setup.}
  \label{fig:Bench}
\end{figure*}

We consider the benchmark problem defined by Class et al.~\cite{Bench1},
which concerned with leakage of injected $\CO$ into the aquifer through a leaky well.
This benchmark is set up based on the studies in
~\cite{Nordbotten2004,Bench11a,Bench11b}.  The focus is on a leakage scenario
consisting of three hydrogeological layers---two aquifers separated by an
aquitard ---that are characterized by uniform thickness and
homogeneous parameters.  The model involves one $\CO$ injection well and one
leaky well.  The leaky well is located at the center of the domain with the
injection well 100~m away.  The domain has lateral dimensions of 1000~m
$\times$ 1000~m.  The sketch in Figure \ref{fig:Bench} summarizes the model geometry
and illustrates a 2D section of the 3D domain. 
The injected $\CO$ spreads within the aquifer and once it
reaches the leaky well, it connects the two aquifers and rises to a shallower
aquifer.  The two aquifers are each 30~m thick and the separating aquitard
(caprock) is 100~m thick. Spatial heterogeneity is considered only through the
different layers according to different geological media.  The formation has a
permeability $K_A$, and the leaky well, which is modelled as a porous medium, 
has a permeability $K_L > K_A$. Changes in fluid properties of $\CO$ are considered in 
this paper.
Note that the $\CO$ and brine fluid
properties (e.g., density and viscosity) depend on the aquifer conditions, the
temperature T, the $\CO$ pressure $p_\nw$, the brine salinity
$s_\salt=0.1$ kg NaCl per kg, and the mass fraction of $\CO$ in brine. 

\subsection{Governing equations}
The physical process of $\CO$ injection in geologic reservoirs, including
solubility trapping, is a non-isothermal two-phase two-component flow in porous
media, which is governed by a system of coupled nonlinear partial differential
equations.  In this model, the water-rich phase (brine, $\w$) and the carbon
dioxide-rich phase ($\CO$, $\nw$) consist of two components (water, $\cw$ and
$\CO$ component, $\cn$), as the solubility of the components in the phases has
to be taken into account.

Local equilibrium phase exchange of the components in the
phases is assumed to hold. Mass balance of the two components 
yields two partial differential equations for the components $\beta$ in the phases $\alpha$
\begin{align}
 & \phi \dt ( \rho_{\mol,\w}  \x_\w ^\cw S_\w  + 
                  \rho_{\mol,\nw} \x_\nw^\cw S_\nw        
           ) 
       + \di ( \rho_{\mol,\w } \x_\w ^\cw \bV_\w +
                  \rho_{\mol,\nw} \x_\nw^\cw \bV_\nw  
           ) + 
       \di ( \bJ_\w^\cw + \bJ_\nw^\cw) = f^\cw, 
       \label{Eq:PhysicalModelForMixture1}\\ 
 &\phi \dt ( \rho_{\mol,\w}  \x_\w ^\cn S_\w  + 
                  \rho_{\mol,\nw} \x_\nw^\cn S_\nw        
            ) +
       \di  ( \rho_{\mol,\w } \x_\w ^\cn \bV_\w +
                  \rho_{\mol,\nw} \x_\nw^\cn \bV_\nw  
            ) + 
       \di  ( \bJ_\w^\cn + \bJ_\nw^\cn ) = f^\cn.
       \label{Eq:PhysicalModelForMixture2}
\end{align}
Here, we denote by $\phi$ the porosity, $\rho_{\mol,\alpha}$
the molar density of phase $\alpha$, $S_\alpha$ the $\alpha$
saturation, $\bV_\alpha$ the velocity phase $\alpha$,
$\bJ_\alpha^{\beta}$ a diffusive flux of the $\beta$ component into
the $\alpha$ phase, $\x_\alpha^\beta$ the molar fraction of component
$\beta$ in phase $\alpha$, $f^{\beta}$ a source term for the
$\beta$ component. 

We also include the energy balance equation for thermal processes that may
occur while the $\CO$ migrates through the formation.  Under the assumption of
local thermal equilibrium, only one energy balance equation for the
fluid-filled porous medium is necessary

\begin{equation}
  \phi \frac{\partial \left( \sum_\alpha \rho_{\alpha}
      u_\alpha S_\alpha \right)}{\partial t} + \left( 1 -
    \phi \right) \frac{\partial \rho_{\text{s}} c_{\text{s}}
    T}{\partial t}  
 - \di \left( \lambda_{\text{pm}} \nabla T \right)
   + \sum\limits_\alpha \di \left\lbrace \rho_{\alpha} h_\alpha
     \bV_\alpha \right\rbrace 
 =  f^h, \quad \alpha\in \{\w,\nw\}, %
\end{equation}
where $\rho_{\alpha}$ is the mass density of phase $\alpha$, $T$
is the temperature, $u_\alpha$ is the specific internal energy, $\rho_s$ and $c_\text{s}$ are the density and 
the  specific heat capacity of the porous medium, respectively, $h_\alpha$ is the  specific enthalpy, $f^h$ is the  heat
source term and $\lambda_\text{pm}$ is the effective heat conductivity of the fluid-filled porous medium.  
The saturation of the $\alpha$ phases and the molar fractions (used to describe the composition of phases) satisfy
\[
  S_\w + S_\nw = 1, \quad  
  \x_\w^\cw + \x_\w^\cn = 1, \quad 
  \x_\nw^\cw + \x_\nw^\cn = 1.
\]

The relation between the phase pressures is given through the capillary pressure using the
Brooks-Corey model  \cite{1964BrooksCorey}:
\begin{align}
 p_\text{cap}(S_\w) = p_\nw - p_\w.
\end{align}

The phase velocities $\bV_\alpha$ are given by the extended Darcy's law for multiphase flow
in porous media:
\[
   \bV_\w  = - \absK \frac{k_{r\w}(S_\w)}{\mu_\w} 
               \left( \nabla p_\w -\rho_{\w}\cdot \g \right), \quad 
   \bV_\nw = - \absK \frac{k_{r\nw}(S_\nw)}{\mu_\nw} 
               \left( \nabla p_\nw -\rho_{\nw}\cdot \g \right),
\]
where $\absK$ denotes the absolute permeability tensor,
$k_{r\alpha}$  denote the relative permeability functions, and $\g$ is the gravity vector.
Following Fick's law, the diffusive flux of component $\beta$ in phase $\alpha$ is given by
\begin{align}\label{eq:flux}
  \bJ_\alpha^\beta = - \bD_\alpha^\beta \rho_{\mol,\alpha} \nabla
  \x_\alpha^\beta,\quad \bJ_\alpha^\cw + \bJ_\alpha^\cn = 0,
\end{align}
where $\bD_\alpha^\beta$ is the diffusion coefficient of component $\beta$ in
phase $\alpha$.

\begin{table}[h] \centering
\caption[]{Fluid properties and simulation parameters.}
\begin{tabular}{ll}
  \hline
  Parameter 
             & Value/function  \\
  \hline
   $\CO$ mass density, $\rho_\nw$      & $f(T,p_\nw)$ \cite{SpanandWagner96} \\
   Brine mass density, $\rho_\w$          & $f(T,p_\w,s_\salt,x_\w^\cn)$  \cite{IAPWS} \\
  $\CO$ viscosity, $\mu_\nw $              & $f(T,p_\nw)$ \cite{Fenghour98} \\
  Brine viscosity, $\mu_\w $                  & $f(T,s_\salt)$ \cite{Batzle92} \\
  $\CO$ enthalpy, $h_\nw$                   & $f(T,p_\nw)$ \cite{SpanandWagner96} \\
  Brine enthalpy, $h_\w$                       & $f(T,p_\w,s_\salt,X_\w^\cn)$ \cite{IAPWS} \\
  Mutual solubilities, $x_\alpha^\beta$  & $f(T,s_\salt,p_\nw)$ \cite{2005SpycherPruess} \\
  Brine salinity, $s_\salt$                         & 0.1 kg NaCl per kg \\
  Residual brine saturation, $S_{r\w}$   & 0.2  \\ 
  Residual $\CO$ saturation,$S_{r\nw}$  & 0.05 \\
  Relative permeability, $k_{r\alpha}$      &  Brooks and Corey \cite{1964BrooksCorey} \\
  Capillary pressure, $p_\text{cap}(S_\w)$            & Brooks and Corey \cite{1964BrooksCorey} \\
  Entry pressure, $P_e$                           &  $10^4$ Pa \\
  Brooks-Corey parameter,  $\lambda$   & 2.0 \\
  Leaky \& injection well radius                 & 0.15 m \\
  \hline
\end{tabular}
   \label{tab:Bench1fluidproperties}  
\end{table}

To close the system, 
the fluid properties of $\CO$ are calculated as functions of pressure and temperature. 
The properties of brine additionally depend on the salinity and on the mole fraction of $\CO$ in brine.
Detailed information on dependencies of the fluid properties is given in Table \ref{tab:Bench1fluidproperties}. 

\subsection{Simulation scenario} 
Since the aquitard is modeled as a layer of impermeable rock, for
computational efficiency, only the aquifers and the leaky well are discretized.
The boundaries between the discretized regions and the aquitard are modeled as
no-flow boundaries.  The initial conditions in the domain include a hydrostatic
pressure distribution that is dependent on the brine density, and a geothermal
temperature distribution that depends on the geothermal gradient.  The
geothermal gradient is taken to be 0.03 \text{K/m}, and the initial temperature
at the bottom (at 3000 m depth) is $100 {}^\degree\text{C}$.  The aquifers are
assumed to be initially saturated with brine. The initial pressure at the
bottom of the domain is taken to be 3.086 $\times$ $10^7$ Pa.  The lateral
boundary conditions are constant Dirichlet conditions and equal to the initial
conditions.  No-flow boundary conditions, for both brine and $\CO$, are assumed
at the top and bottom of the domain.

In the benchmark setup, $\CO$ is being injected at a constant rate of 8.87
\text{kg/s};  this corresponds to 1600 \text{m}${}^3$ per day at reservoir
conditions. The total simulation time is 1500 days. All relevant parameters
used for the simulation are given in Table \ref{tab:Bench1fluidproperties}. For
more details we refer the reader to \cite{Bench1}.  The $\CO$ leakage rate (denoted by
$\Qleak$ in this paper), which is the output quantity of interest (QoI) of the
benchmark study, is defined as the total $\CO$ mass flow at midway between the
top and bottom aquifers divided by the injection rate, in percent. 

We utilize the DuMuX simulator \cite{Dumux} to solve
\eqref{Eq:PhysicalModelForMixture1}--\eqref{eq:flux}. For spatial
discretization, we use the so-called Box method \cite{Huber2000}, which is a
vertex-centered finite volume method. For temporal discretization,  we use
implicit Euler.

%
%
\newcommand{\xij}{\xxi^{(j)}}
\newcommand{\Omegas}{\Theta}
\section{Background on spectral methods for uncertainty quantification}
\label{sec:spectralUQ}
We begin our discussion of spectral UQ
methods, by using the problem of uncertainty quantification for $\CO$ leakage,
which we denote by  $\Qleak = \Qleak(t, \vec{q})$ where  
$t$ denotes time. The vector $\vec{q}$ contains a set of
parameters defining various physical properties of the system. These parameters
appear as coefficients, or boundary or volume forcing terms in the governing PDEs.
The elements of this vector are, in
practice, known only approximatively and are hence considered uncertain.  An important
consideration in obtaining high-fidelity predictions is to quantify the impact
of these parametric uncertainties on the model observables such as $\Qleak$.  
To this end, we
model the uncertain parameters as random variables that are parameterized by a
vector $\xxi$ of %
canonical random variables.  Hence, 
we will have $\Qleak = \Qleak(t, \vec{q}(\xxi))$.  The random vector
$\xxi$ fully characterizes the uncertain parameter vector $\vec{q}$ and,
therefore, we can unambiguously use the simpler notation $\Qleak(t,
\xxi)$ for the uncertain $\CO$ leakage, a convention which we follow for the
other uncertain model variables below as well.

Below we seek to approximate the nonlinear mapping $\xxi \mapsto \Qleak(t, \xxi)$ through
a spectral representation of the form
\[
    \Qleak(t, \xxi) = \sum_{k = 0}^\infty c_k(t) \Psi_k(\xxi),
\]
where the $\Psi_k$'s form an orthogonal basis in an appropriate Hilbert space
(discussed below), and $c_k(t)$ are expansion coefficients.  Such
a spectral representation can then be used as a cheap-to-evaluate
\emph{surrogate} for the parameter-to-observable map, $\xxi \mapsto
\Qleak(t, \xxi)$. This enables efficient methods for characterizing
the uncertainties in $\Qleak$ that replace expensive PDE solves
by cheap evaluations of the surrogate. 

\subsection{Notation and definitions}
We denote by $\left( \Omega, \mF,  \pro \right)$ 
a probability space, where $\Omega$ is the sample space, $\mF$ is an
appropriate $\sigma$-algebra on $\Omega$, and $\pro$ is a probability measure.
For a random variable $\xi$ on $\Omega$, we write $\xi \sim \mU(a,b)$ to mean
that $\xi$ is uniformly distributed on the interval $\left[a,b\right]$ and  $\xi
\sim \mN(0,1)$ to mean that $\xi$ is a standard normal random variable.
We use the term iid for a collection of
random variables to mean that they are independent and identically distributed.
The distribution function of a random variable $\xi$ on $\Omega$ is given by
$F_\xi(x) = \pro(\xi \le x)$ for $x\in \R$.

In the present work, we consider models with finitely many uncertain
parameters. We parameterize these uncertain parameters by a finite collection
of real-valued independent random variables $\xi_1,\cdots,\xi_\dim$ that are defined on $\Omega$. 
We let $F_\xxi$ denote the joint distribution function of the
random vector $\xxi = \left( \xi_1, \cdots, \xi_\dim \right)^T$. Since the
$\xi_i$ are independent, $F_\xxi(x) = \prod_{i=1}^{\dim} F_i(x_i)$ for $x \in \R^\dim$,
where $F_i$ is the distribution function of the $i$th coordinate. 

The random vector $\xxi$ takes values in $\R^\dim$.
In fact, it is sufficient to consider the subset $\Omegas$ of $\R^\dim$
given by the support of the distribution function $F_\xxi$.
Following common practice, we work in the image probability space $\left(
\Omegas,\mB(\Omegas),F_\xxi\right)$, where $\mB(\Omegas)$ is the Borel
$\sigma$-algebra on $\Omegas$ (which is a standard choice). For 
notational convenience we suppress $\mB(\Omegas)$ below and denote the image probability space
by $(\Omegas, F_\xxi)$. We denote the expectation of a random variable
$X : \Omegas \to \R$ by 
\[\langle X  \rangle = \int_{\Omegas} X(\vec{s}) F_\xxi(\dd\vec{s}).\] 
The space of square-integrable random variables on $\Omegas$,
$\tL^2(\Omegas,F_\xxi)$, is endowed with the inner product $(\cdot,\cdot)$
defined by $(X,Y) = \int_{\Omegas} X(\vec{s}) Y(\vec{s})
F_\xxi(\dd\vec{s}) = \langle XY \rangle$, and the corresponding induced norm 
$\| \cdot \| = (\cdot,\cdot)^{1/2}$.

\subsection{Polynomial chaos expansion}
In the case $\xi_i \overset{\text{iid}}{\sim} \mN(0,1)$, for $i = 1, \ldots, \dim$, 
any $X \in \tL^2(\Omegas,F_\xxi)$ admits an expansion of the form,
\begin{align}\label{eq:ExpansionForm}
X = \sum_{k=0}^{\infty} c_k \Psi_k,
\end{align}
where $\{\Psi_k\}_0^{\infty}$ is a complete orthogonal set consisting of
$\dim$-variate Hermite polynomials, and the series converges in
$\tL^2(\Omegas,F_\xxi)$.  The infinite series representation of $X$ is known as
the polynomial chaos (PC) or Wiener--Hermite expansion of
$X$~\cite{Wiener:1938,Cameron:1947,Ghanem:1991a,knio:2010}.  The Wiener-Hermite
expansion is the appropriate choice in the case the model parameters are
parameterized by normally distributed random variables. In the case where the
sources of uncertainty follow other distributions, alternative
parameterizations and polynomial bases are adopted %
~\cite{knio:2010}.  For example, in the case where $\xi_i
\overset{\text{iid}}{\sim} \mU(-1,1)$ the appropriate PC basis is given by the
$d$-variate Legendre polynomials.

\paragraph{Tensor product construction of a multivariate PC basis}
Let $\xxi = (\xi_1, \ldots, \xi_\dim)$, where  
$\xi_i$ are independent random variables that are distributed according to
common choices given by standard normal, uniform, or beta distributions.
We work with a multivariate PC basis that is obtained through a tensor product of appropriate
one-dimensional bases.  More precisely, if we denote by $\{\psi_j(\xi_i)\}_{j =
1}^\infty$ the one-dimensional orthogonal polynomial basis corresponding to
$\xi_i$ (with the choice of basis dictated by the distribution of $\xi_i$), we
form the multivariate PC basis $\{\Psi_k\}_{k = 0}^\infty$ as follows: 
\begin{equation}\label{equ:multivariatePC}
	\Psi_k(\xxi) = \prod_{i=1}^{\dim} \psi_{\alpha_i^k}(\xi_i), \qquad \xxi \in \Omegas,
\end{equation}
where $\alpha^k = \left( \alpha_1^k, \alpha_2^k, \cdots, \alpha_d^k\right)$ is
a multi-index, and $\alpha_i^k$ indicates the order of the 1D polynomials in
$\xi_i$.  For example, if $\xi_i$ is standard normal, then $\psi_{\alpha_i^k}$
is the Hermite polynomial of order $\alpha_i^k$.
With this basis, any $X \in L^2(\Omegas, F_\xxi)$ admits an expansion of the form:
$
X = \sum_{k=0}^{\infty} c_k \Psi_k
$, which is known as the generalized polynomial chaos expansion of $X$. 
In computer implementations, we will approximate $X(\xxi)$ with a truncated series,
\begin{align}\label{eq:FiniteExpansion}
	X(\xxi) \approx \sum_{k=0}^{P} c_k \Psi_k(\xxi)
\end{align}
where $P$ is specified based on the choice of truncation strategy. In the
present work, we consider truncations based on the total degree of the
polynomials in the series. In this case, letting $p$ be the largest (total)
polynomial degree allowed in the expansion, it is straightforward to show that
$P = (\dim+p)!/(\dim!\,p!)-1$, where as before $\dim$ is the dimension of the
uncertain parameter vector $\xxi$.

Note that with $X$ expanded as in~\eqref{eq:FiniteExpansion}, using the orthogonality
of the basis $\{\Psi_k\}_0^P$ and the convention that $\Psi_0=1$, we have 
immediate access to its first and second moments:
\[
\langle X \rangle    = c_0, \quad \langle X^2 \rangle  = \sum_{k=0}^P c_k^2 \langle \psi_k^2 \rangle,
\]
from which we also get $\var{X} = \sum_{k=1}^P c_k^2 \langle \psi_k^2 \rangle$.

\paragraph{Tests of convergence}
To assess accuracy of a PC expansion, one could begin by studying convergence in 
distribution. A practical method of doing this is by tracking the probability density 
function (pdf) of the PC expansion (which can be approximated efficiently by sampling
the expansion and using for example a Kernel Density Estimation (KDE) method)
as the order of the expansion is increased.
Moreover, 
to get further confidence in the spectral representation of a random variable
$X(\xxi)$, one can use the relative $L^2$ error, $E_\text{rel}$, between $X$ and its truncated
PC representation: 
\begin{equation}\label{sec:L2_error}
E_\text{rel}^2 := \frac
{\displaystyle\int_{\Omegas} | X(\vec{s})-\sum_{k=0}^{P} X_k \Psi_k(\vec{s})|^2 F_\xi(\dd \vec{s}) } 
{\displaystyle \int_{\Omegas} |X(\vec{s})|^2 F_\xi(\dd\vec{s})},
\end{equation}
which can be approximated using either quadrature or sample averaging.

\subsection{Non-intrusive spectral projection}
Let $X$ belong to $\tL^2(\Omegas,F_\xxi)$. As mentioned in the introduction,
non-intrusive methods aim at computing the PC coefficients in the finite expansion 
\eqref{eq:FiniteExpansion} via a set of deterministic evaluations of $X(\xxi)$ for specific 
realizations of $\xxi$. Observe that since $\{\Psi\}_0^P$ form an orthogonal system, we have:
$(X,\Psi_k) = \Big(\sum_{l=0}^P c_l \Psi_l, \Psi_k \Big) 
		  = \sum_{l=0}^P c_l (\Psi_l, \Psi_k)
		  = c_k (\Psi_k,\Psi_k)$,
so that the coefficient $c_k$ is given by 
\[
c_k = \frac{\langle X \, \Psi_k \rangle}{\langle \Psi_k^2 \rangle}.
\]

The moments $\langle \Psi_k^2 \rangle$ of known orthogonal polynomials  
can be computed analytically, and hence, the determination of coefficients $c_k$ amounts to the 
evaluation of the moments $\langle X \Psi_k \rangle$. 
In the non-intrusive spectral projection (NISP) approach, these moments are approximated via quadrature: 
\begin{align}\label{eq:Integration}
        \langle X \Psi_k \rangle = 
	\int_{\Omegas} X(\vec{s}) \Psi_k(\vec{s}) F_\xxi(\dd\vec{s})
        \approx \sum_{j=1}^{N_q}  \om_j X(\xij) \Psi_k(\xij),
\end{align}
where $\xij \in \Omegas$ and $\om_j$ are the nodes and weights of an appropriate quadrature formula.
Note that in this formulation, the same set of nodes is used to compute all coefficients $c_k$. Hence, 
the complexity of NISP, measured in the number of evaluations of $X(\xxi)$ (i.e., the number of model solves), 
scales with the number of quadrature nodes $N_q$.
These multi-dimensional quadrature rules are constructed by 
full or partial tensorization of one-dimensional quadrature formulas. 
Therefore, the number $N_q$ of quadrature nodes scales with the dimension of the 
uncertain parameter---a phenomenon commonly referred to as the curse of dimensionality.
In the present work, we work with a small number of uncertain parameters, and hence a 
full-tensor Gaussian quadrature was found feasible. However, for higher-dimensional
problems, sparse grids, or adaptive sparse grids are more 
suitable~\cite{BlatmanSudret11,knio:2010,WinokurConradSrajEtAl13}.

We remark that the efficient construction of PC expansions via non-intrusive methods
has resulted in significant research activity in recent years. The efforts include
adaptive pseudo-spectral projections~\cite{ConradMarzouk13,BryantSergeWildey} as well as regression-based approaches
that incorporate sparsifying penalty methods~\cite{HamptonDoostan15,YanGuoXiu12,PengHamptonDoostan}. While the goal of the
present work is not the study of such methods, nor their extensions, we point them out 
as potential solutions for the problems with higher-dimensional parameters, where one
seeks to utilize PC expansions for uncertainty analysis.

\subsection{Variance-based sensitivity analysis} \label{sec:SA}
An important step in quantifying the impact of parametric uncertainties on
the response of an uncertain system is that of parametric sensitivity analysis. 
In particular, global or variance-based sensitivity 
analysis~\cite{Sobol:1990,Homma:1996,Sobol:2001,Saltelli:2002} 
enable the characterization of the contribution of the individual uncertain parameters
or their interactions to the total variance of the model response. In this section we
outline the concepts from variance-based sensitivity analysis that are used in the present work. 

Consider a square-integrable random variable $X(\xxi)$. 
The first-order (or main effect) sensitivity indices quantify
the effect of the $i^{\text{th}}$ coordinate $\xi_i$ \emph{alone} on the variance of the random variable $X(\xxi)$.
These first-order indices, which we denote by $\mathrm{S}_i$, are defined as follows,
\begin{equation}\label{equ:Si}
   \mathrm{S}_i = \frac{\var{\avemu{ X(\xxi) \myvert \xi_i}}}{\var{X(\xxi)}}, \quad
i \in \{1, \ldots, d\}. 
\end{equation}
Here $\avemu{ X(\xxi) \myvert \xi_i}$ denotes the conditional expectation~\cite{Williams} of $X(\xxi)$ given $\xi_i$.
While the mathematical definition of the first-order indices (and higher-order indices discussed below) 
are given in terms of conditional expectations,
whose numerical approximations involve expensive sampling %
(see e.g.,~\cite{Sobol:2001}), their computation via PC expansion is
straightforward and very
efficient~\cite{Crestaux:2009,Sudret:2008,alexanderian2012global,alexanderian2013spectral}.

We also point out 
the second-order sensitivity indices that describe joint effects. Specifically, for $i, j \in \{1, \ldots, \dim\}$,
we denote by $\mathrm{S}_{ij}$ the sensitivity index that quantifies the contribution of the \emph{interaction} between
$\xi_i$ and $\xi_j$ to the total variance. The mathematical definition of $\mathrm{S}_{ij}$ is as follows,
\begin{equation}\label{equ:Sij}
   \mathrm{S}_{ij} = 
   \frac
   {\mathrm{var}\big\{\avemu{ X(\xxi) \myvert \xi_i, \xi_j}\!\!\big\}}
   {\var{X(\xxi)}} 
   - (\mathrm{S}_i + \mathrm{S}_j). 
\end{equation}
Higher-order joint sensitivity indices (for example $\mathrm{S}_{ijk}$) can be defined
also, but usually are not used in applications. However, in our numerical
computations below we will discuss a sensitivity index, which we call the
\emph{mixed index}, that quantifies the contribution of \emph{all} interactions
among uncertain parameters.  

Another useful variance based sensitivity measure is the \emph{total sensitivity 
index}~\cite{Homma:1996,Saltelli:2002}. The total sensitivity index due to $\xi_i$ is defined by,
\begin{equation}\label{equ:Ti}
T_i
=\frac{\var{X(\xxi)} - \var{\avemu{X(\xxi) \myvert \xxii}}}{\var{X(\xxi)}}
\end{equation}
where $\xxii$ denotes the random vector $\xxi = (\xi_1, \ldots, \xi_d)$ with $\xi_i$ removed:
$\xxii := (\xi_1, \ldots, \xi_{i-1}, \xi_{i+1}, \ldots, \xi_d)$.
Notice that the numerator in~\eqref{equ:Ti}
is the total variance minus the variance of the conditional expectation
$\avemu{X(\xxi) \myvert \xxii}$.
Thus, $T_i$ is the total contribution of $\xi_i$, by itself and
through its interactions with other coordinates, to the variance.

%
%
\section{Uncertain parameters and quantities of interest}\label{sec:UQsetup}
In the present study, we study the effect of uncertainties in reservoir
porosity $\phi$, reservoir absolute permeability $K_A$ and permeability of the leakage well
$K_L$ on the model response. %
To support decision-making based on the
approach presented here, we also consider one design parameter, the $\CO$
injection rate $Q_{\CO}$. This will help to study the influence of the injection rate on
the $\CO$ leakage rate. As reflected in Table~\ref{tbl:params} all uncertain 
parameters are modeled as log-normal. The distributions were adopted based on the setup 
in~\cite{Class2009,Oladyshkin2011}.

\begin{table}[ht]\centering
\caption{Distributions of the uncertain parameters.}
\begin{tabular}{l|l}
Parameter & Distribution\\
\hline
log-porosity              &   $\mathcal{N}(-1.8971, 0.2^2)$\\
log-absolute permeability &   $\mathcal{N}(-30.002, 1.2^2)$\\
log-leaky well permeability  &   $\mathcal{N}(-27.631, 0.3679)$\\
log-injection rate        &   $\mathcal{N}(2.1827, 0.2^2)$
\end{tabular}
\label{tbl:params}
\end{table}

In the analysis below, we focus on the following
model observables that characterize the flow: (a) the $\CO$ leakage through the leaky
well as a function of time,  (b) arrival time of the $\CO$ plume at the leaky
well, (c) the maximum leakage ratio, and (d) the corresponding time;  
these quantities are denoted, respectively, by
$\Qleak$, $\Tarrival$, $\Qmax$, and $\Tmaxleak$.  Note that here the $\CO$
leakage rate is defined as in the benchmark study as the $\CO$ mass flux, in
percent, at midway between  top  bottom aquifer divided by the injection rate;
also, $\Qmax(\xi) = \max_t \Qleak(t, \xi)$.
We also aim to understand the effect of model
uncertainties on the spatially distributed pressure and saturation as functions
of time.  

%
%
\section{Analysis of uncertainties in $\CO$ storage}\label{sec:numerics} 
In our computations, we used NISP based on a fully tensorized Gauss-Hermite
quadrature to compute the spectral expansion of the model output in the PC
basis. To enable a systematic analysis of convergence of the PC expansions, 
we used a hierarchy of quadrature grids. Namely, we constructed full tensor quadrature
formulas using two, three, four, and five nodes in each stochastic dimension, resulting
in non-nested quadrature grids of $n_q^4$ nodes, with $n_q \in \{2, 3, 4, 5\}$. 
The required model evaluations were run on a 20-core Intel Xeon E5-2680
v2 (2.80GHz) workstation. The computational time for a single simulation run
was about 18 hours using five cores. The highest resolution grid of $5^4 = 625$ nodes
supports a fourth-order PC expansion, which was found to provide sufficient
accuracy for the statistical tests needed in our computations.

\subsection{Analyzing uncertain response of $\CO$ leakage}\label{sec:prelim}
\begin{figure}\centering
    \includegraphics[width=0.35\textwidth]{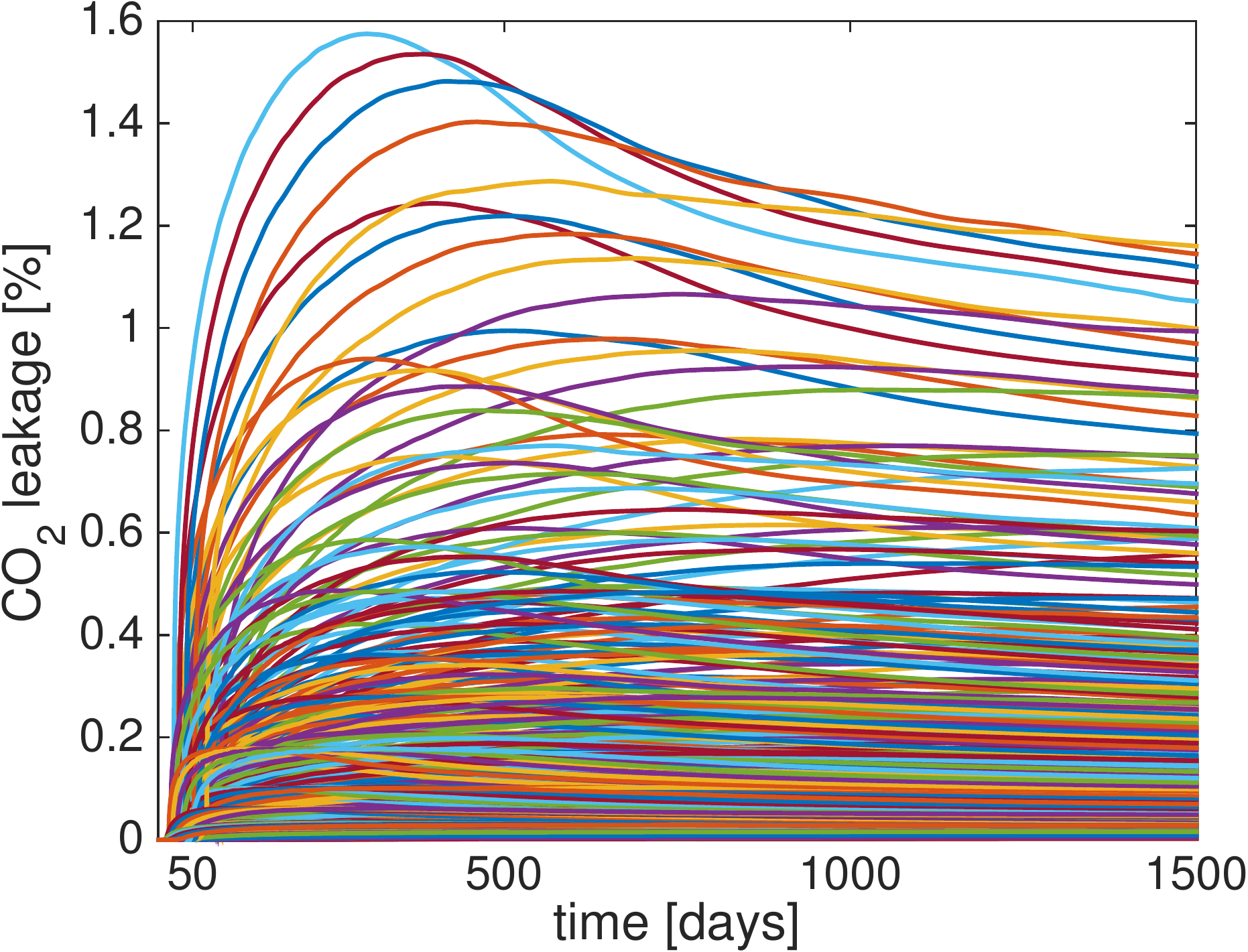}
    \includegraphics[width=0.35\textwidth]{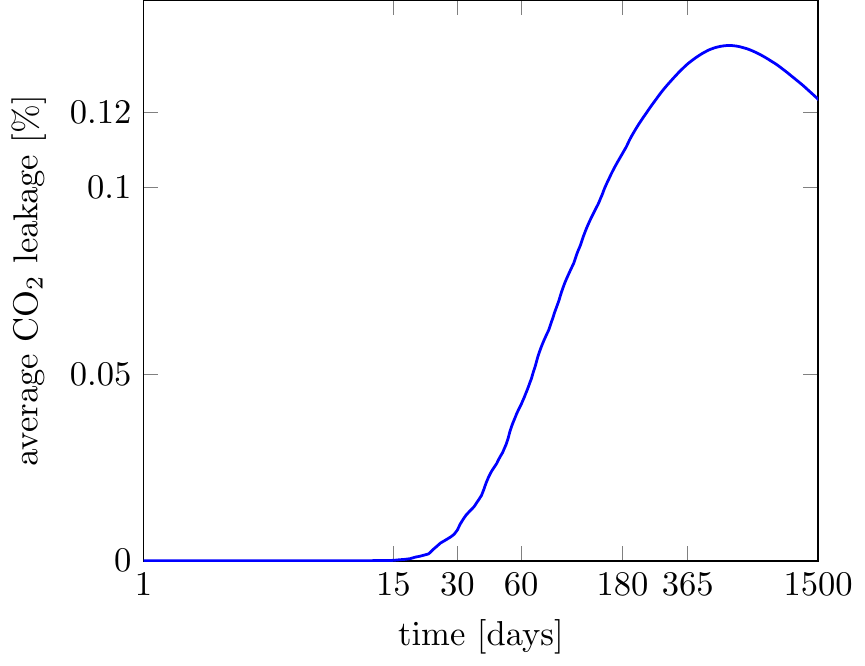} 
    \caption{Evoluation of $\CO$ leakage rate. 
             Left: the curves depict the 625 realizations corresponding to the 
             quadrature points in the uncertain parameter space.
             Right: sample mean of the $\CO$ leakage rate over 
             time (averaged over realizations computed at quadrature nodes).}
    \label{fig:realizations}
\end{figure}

In Figure~\ref{fig:realizations}~(left), we plot the realizations 
of $\Qleak$
as a function of time.
These realizations are obtained by 625 model solves with the parameter
values set according to the 625 nodes of the Gauss-Hermite quadrature.
To understand the solution behavior better, in
Figure~\ref{fig:realizations}~(right) we report the sample mean of the
$\CO$ realization with the averaging done over the realizations
computed at the quadrature points. Note that in that figure, we have
used a log scale for the horizontal axis to provide a clearer picture
of the dynamics of $\CO$ at the early times. 

The results reported in Figure~\ref{fig:realizations} merely
provide an initial screening.  While a small Monte Carlo sample (in the order
of the number of the chosen quadrature nodes) might be used for such an initial
screening,  the model evaluations at the quadrature nodes enable construction
of PC representations for the observables in the expensive-to-simulate
numerical model under study.  It allows for a complete and reliable
characterization of statistical properties of the model observables. In
particular, the PC representations can be cheaply sampled, as many times as
needed, in statistical studies.

\begin{figure}[ht]\centering
    \includegraphics[width=0.335\textwidth]{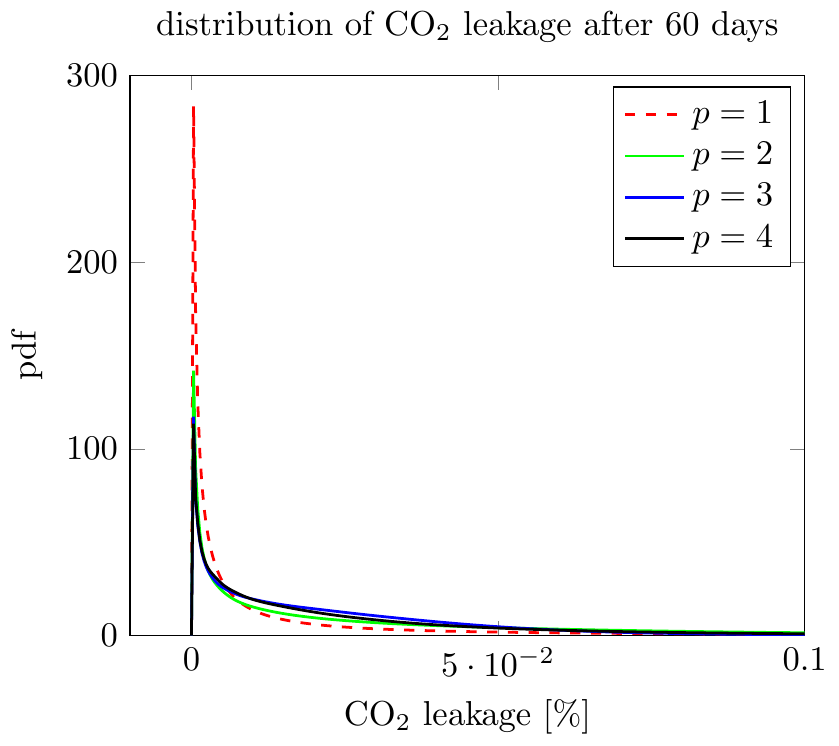} 
    \includegraphics[width=0.3\textwidth]{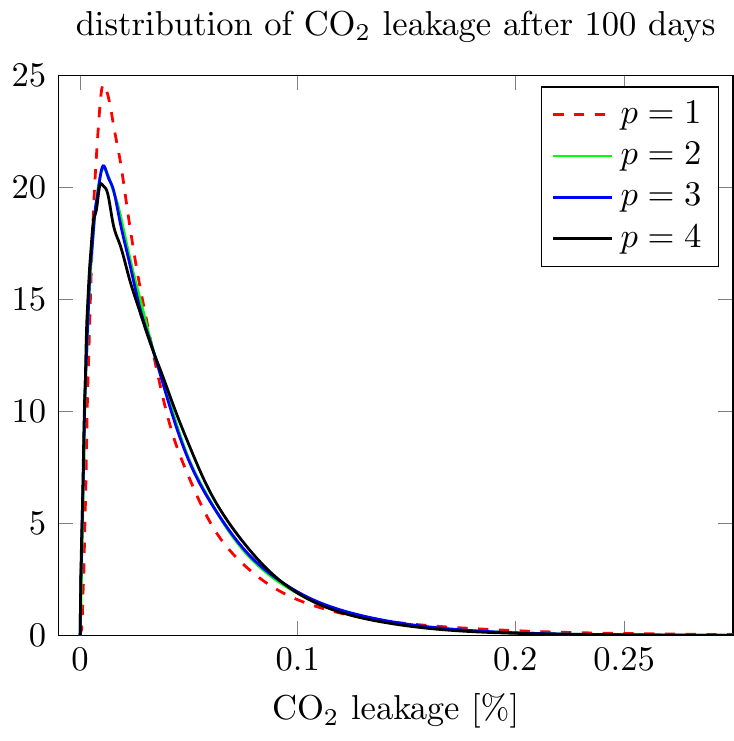}\\ 
    \includegraphics[width=0.325\textwidth]{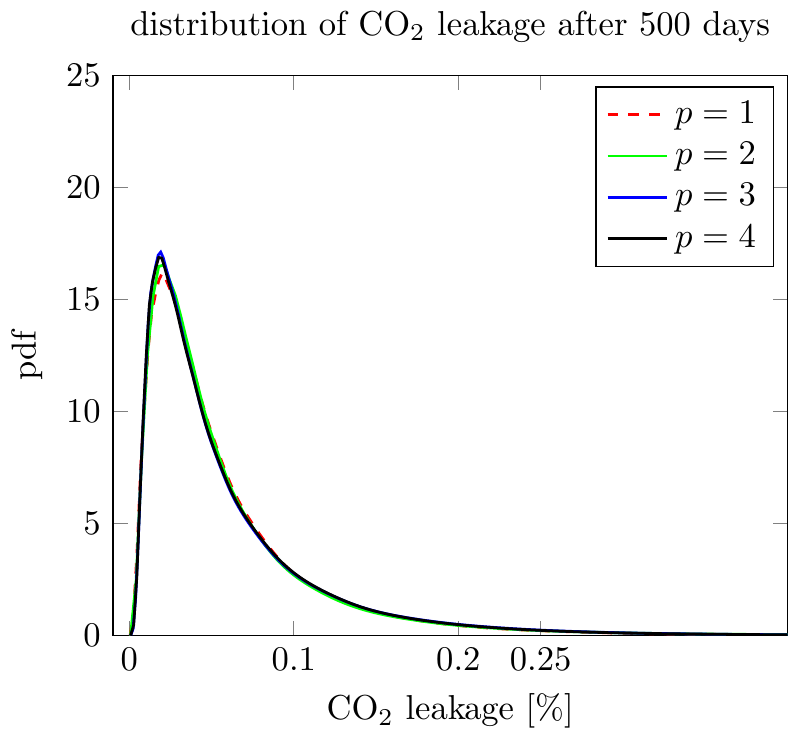} 
    \includegraphics[width=0.3\textwidth]{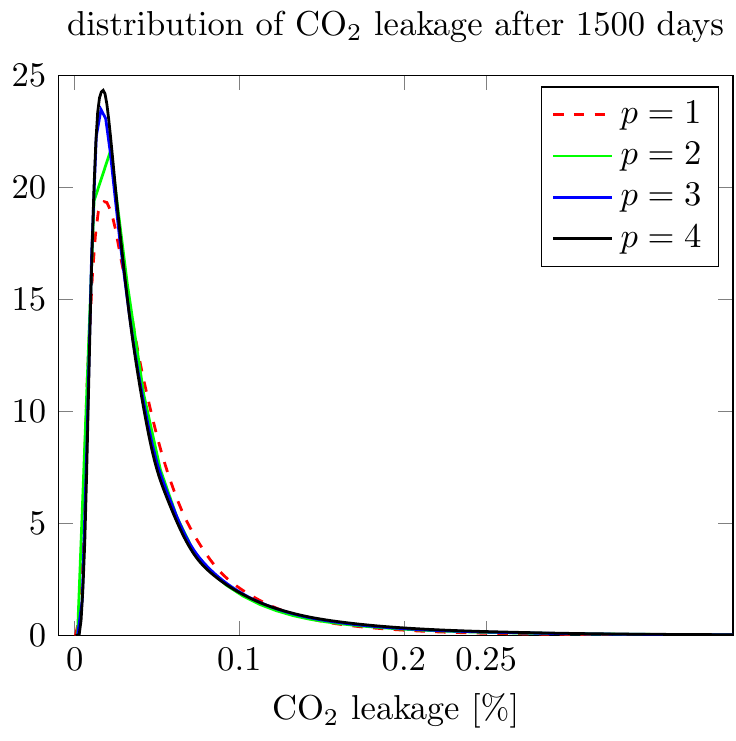} 
\caption{Distribution of $\CO$ leakage at selected times. 
 		$p$ denotes the highest 
		polynomial degree in the truncated expansion. In each 
	        case, the PC expansion was sampled $10^6$ times to 
	        generate the distribution curve.}
\label{fig:conv_pc}
\end{figure}

To obtain a PC representation of $\CO$ leakage, we first project the
log of $\CO$ leakage in a PC basis,
\[
   \log \Qleak(t, \xxi) \approx \sum_{k = 0}^P c_k(t) \Psi_k(\xxi).
\]
The response surface for $\Qleak$ can then be constructed
using
\begin{equation}\label{equ:PCEQleak}
    \Qleak(t, \xxi) \approx \exp\left( \sum_{k = 0}^P c_k(t) \Psi_k(\xxi) \right).
\end{equation}
This log-projection, in particular, ensures the positivity of $\Qleak$.
Figure~\ref{fig:conv_pc} shows instantaneous distributions of $\CO$
leakage rate. These distributions are obtained by sampling the PC-based surrogate
model~\eqref{equ:PCEQleak} at selected times. 
As seen in the plots, the distributions seem to level off as the PC order
is increased to $p=4$ suggesting that a fourth-order expansion is
sufficient.

\begin{figure}[ht]\centering
    \includegraphics[width=0.8\textwidth]{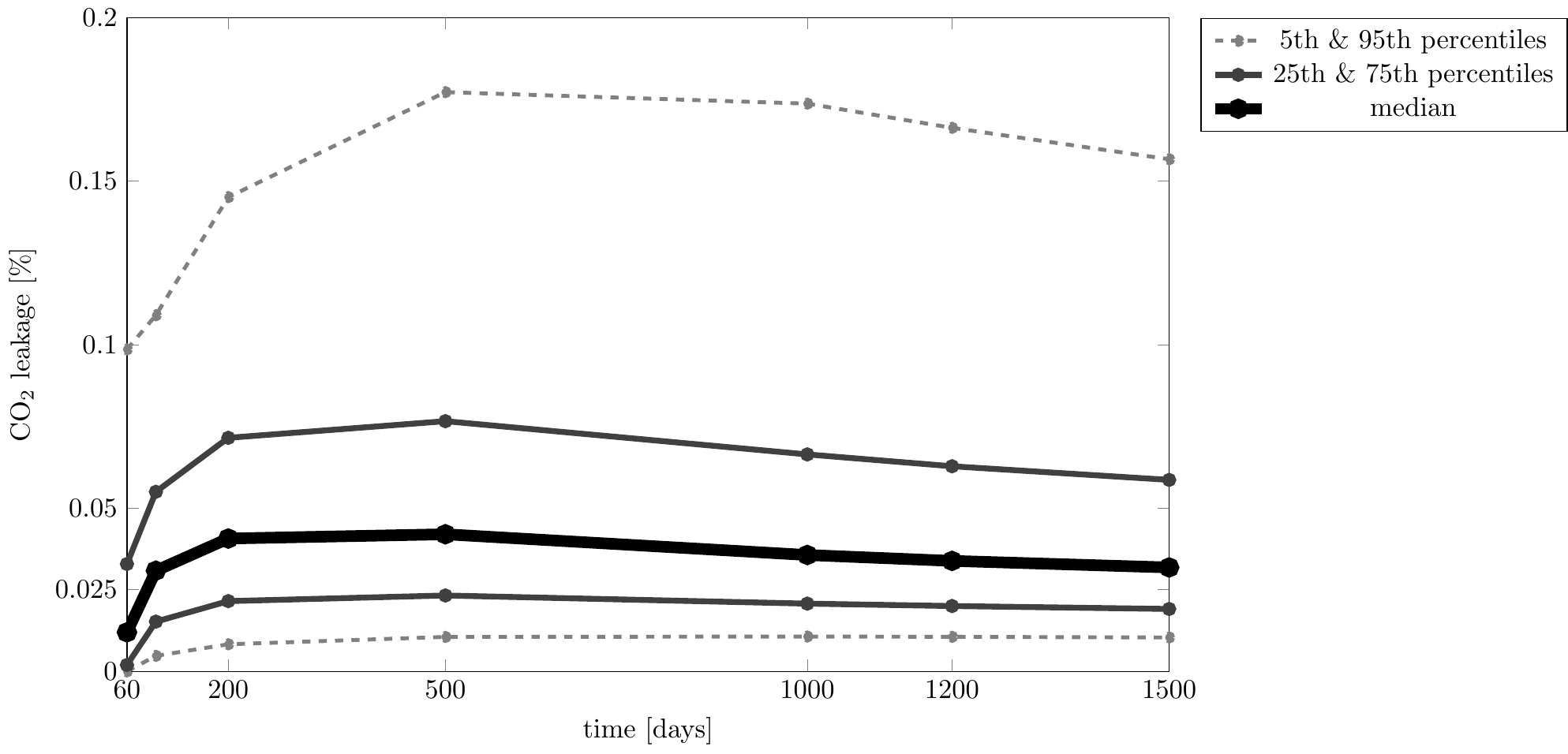} 
    \caption{Percentiles of the distribution of $\CO$ leakage over
      time.} 
    \label{fig:boxplots}
\end{figure}

To further illustrate the evolution of the distribution of $\CO$
leakage over time, we show in Figure~\ref{fig:boxplots} 
$p$-percentiles of the distribution at different times for 
$p \in \{5, 25, 50, 75, 95\}$. 
These plots are generated by
sampling the PC representation of $\CO$ leakage with a Monte Carlo
sample size of $10^6$.  These results further illustrate the skewed
distribution of $\CO$ leakage and its spread. 

\begin{figure}
    \begin{tabular}{lll}
    \includegraphics[width=0.31\textwidth]{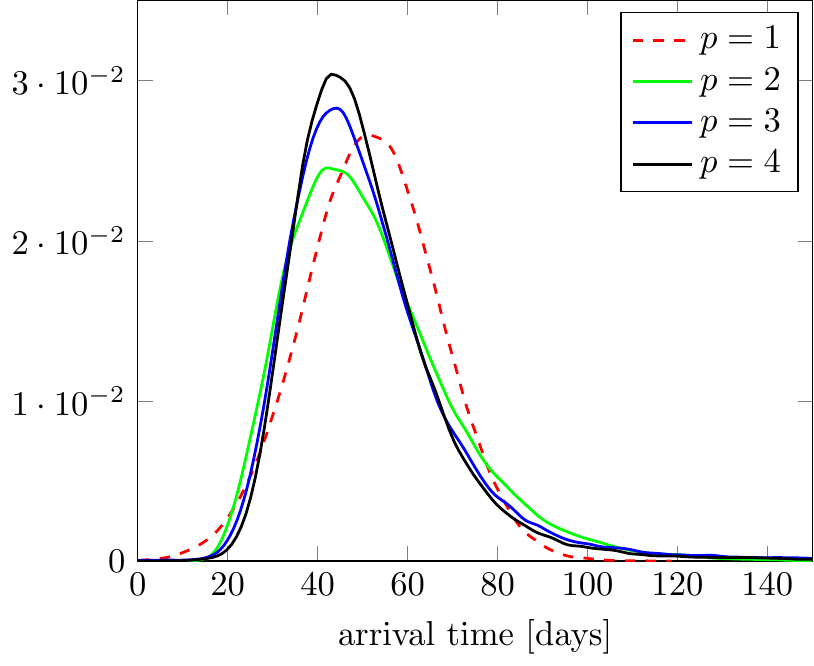} & 
    \includegraphics[width=0.29\textwidth]{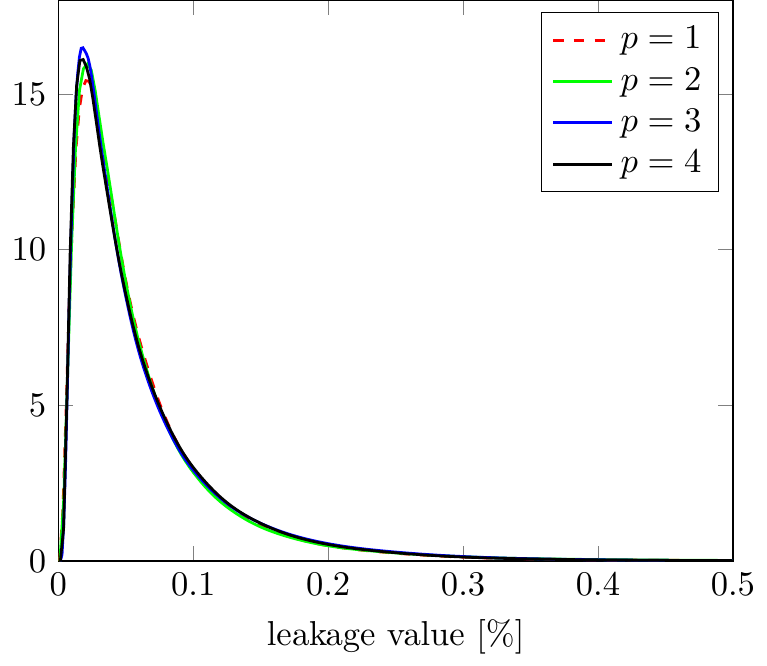} & 
    \includegraphics[width=0.325\textwidth]{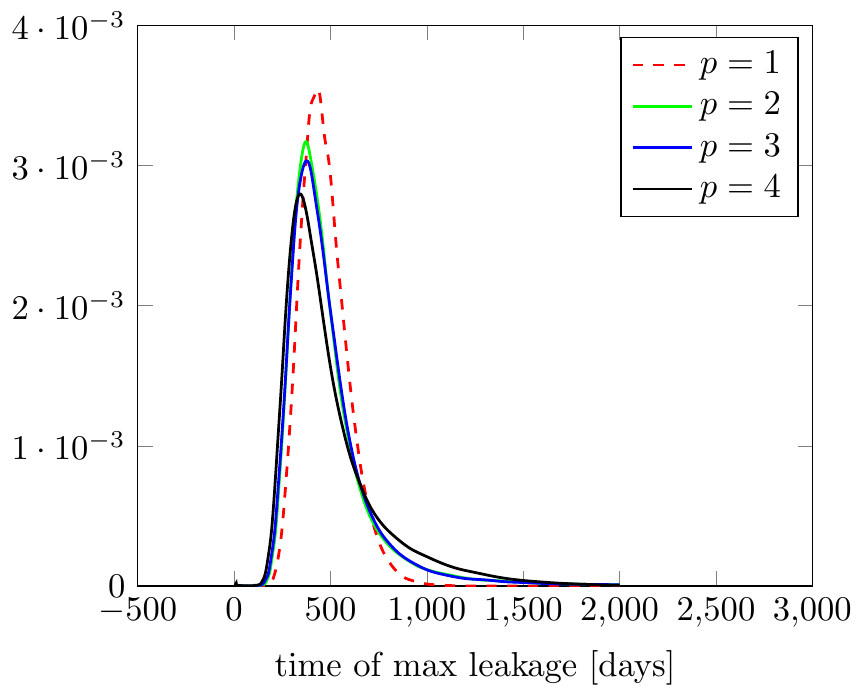}    
    \end{tabular}
    \caption{Distribution of the arrival time of the $\CO$ plume at the leaky well, 
    		the maximum leakage ratio and the corresponding time of the maximum 
		leakage ratio.
 		$p$ in the legend denotes the highest 
		polynomial degree in the truncated expansion. In each 
	        case, the PC expansion was sampled $10^6$ times to 
	        generate the distribution curve.}
    \label{fig:arriv}
\end{figure}

To get a more complete picture of the response of the model to
parametric uncertainties, we use spectral representations to
approximate the (uncertain) arrival time of the $\CO$ plume at the
leaky well (which is defined as the time at which the leakage value is
greater than $3.0 \times 10^{-3}\%$), the maximum leakage ratio and
the corresponding time; see Figure \ref{fig:arriv}.

The use of nonlinear relative permeability-saturation relation, as done in this
work, leads to a later arrival time compared to the case of linear relative
permeability~\cite{Bench11b,Bench1}.  The reason behind this is that using the
nonlinear relation, the sum of the relative permeability values of the brine
and $\CO$ phases is less than unity for most saturations; this leads to an
overall reduced mobility of the flow and thus to a reduced leakage at early
times with later arrival and lower peak as the leakage rate does not rise as
high as in the case of linear relative permeability~\cite{Bench11b,Bench1}.
This later arrival time is seen in our numerical results, for example by
looking at the expected value of the arrival time, easily obtained using 
the PC representation of $\Tarrival$, that is approximately 52
days.  This should be contrasted with the arrival reported in~\cite{Bench1},
where averaging the arrival times computed using different numerical solvers
that use the linear relative permeability-saturation relation is about 9 days. 

\begin{figure}[ht]
\centering
\includegraphics[width=0.55\textwidth]{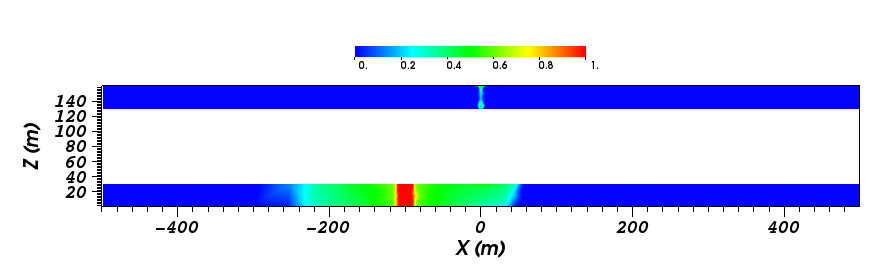}
\includegraphics[width=0.55\textwidth]{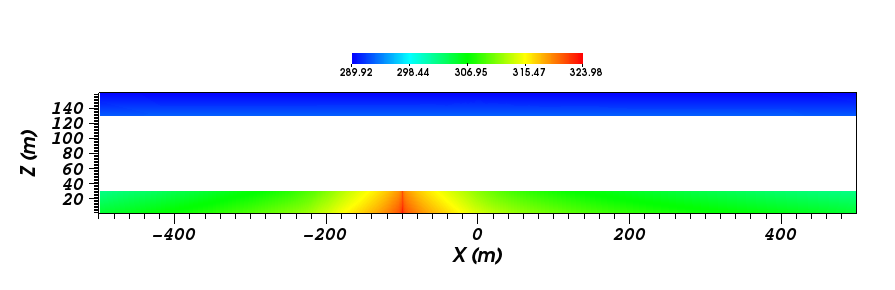}
\caption{A typical realization of $\CO$ saturation (top), and a typical realization of
pressure (bottom) at $120$ days.}
\label{fig:2dslide_sat_t120}
\end{figure}

There are further effects contributing to the later arrival time of
$\CO$ at the leaky well, like increased influence of the viscous
forces in the system due to the lower relative permeabilities,
compared to buoyancy due to density differences, which makes the shape
of the plume become more cylindrical \cite{Class2009}; see Figure
\ref{fig:2dslide_sat_t120} (top) that illustrates the saturation of $\CO$
after 120 days, obtained for one realization model. Figure
\ref{fig:2dslide_sat_t120} corresponds to a vertical slice through the
middle of the domain. In that figure (bottom image), we also show a typical realization 
of the pressure field along the same vertical slice.

\begin{figure}[ht]
\centering
\includegraphics[width=0.35\textwidth]{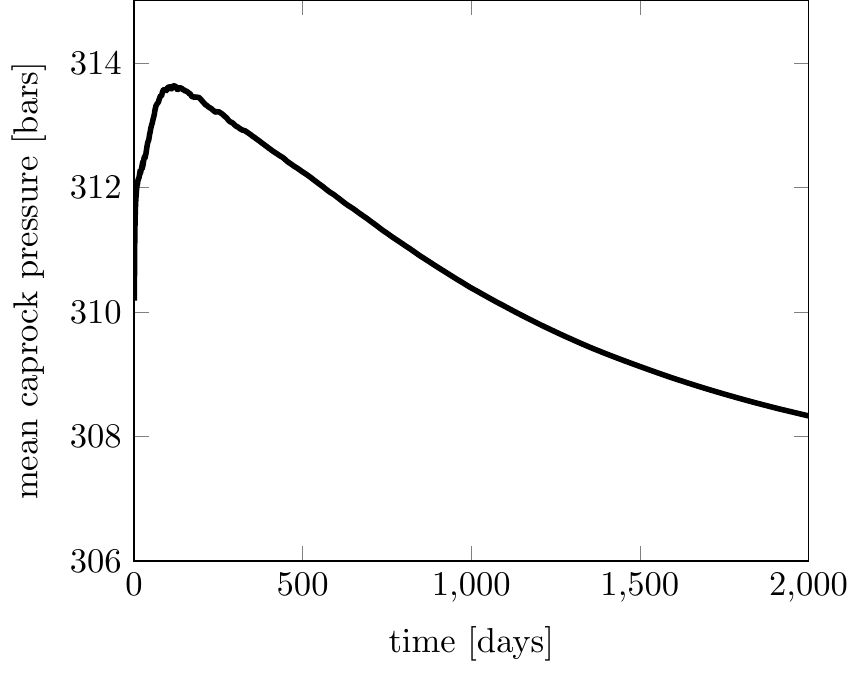}
\caption{Mean caprock pressure as a function of time.}
\label{fig:meancaprock}
\end{figure}

As a result of the increased overall resistance to the flow, the
leakage rate rises smoothly when the $\CO$ reaches the well and then
approaches steady-state. 
This behavior can be attributed to the
lateral boundary conditions that influence the pressure in the domain.
In Figure \ref{fig:meancaprock}, we report the time evolution of the
mean caprock pressure, where we see a reduced pressure over time. This
is the reason for the leakage rate to start decreasing after the peak
of the arrival of the $\CO$ flux at the leaky well~\cite{Bench11b}.

\subsection{Accuracy of the PC representation for $\CO$ leakage}
\begin{figure}[ht]
  \centering
    \includegraphics[width=0.32\textwidth]{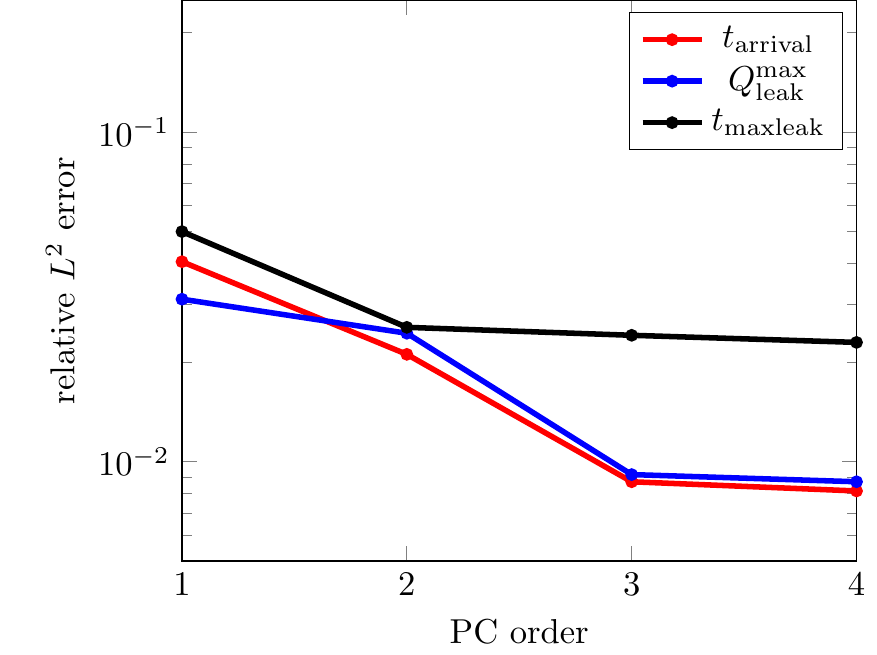}
    \includegraphics[width=0.3125\textwidth]{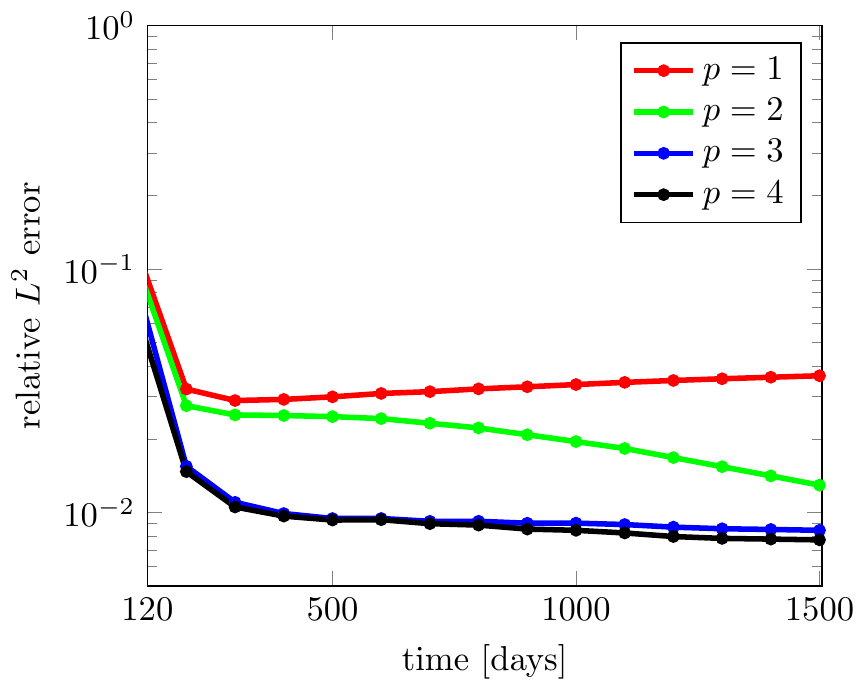}
    \includegraphics[width=0.3\textwidth]{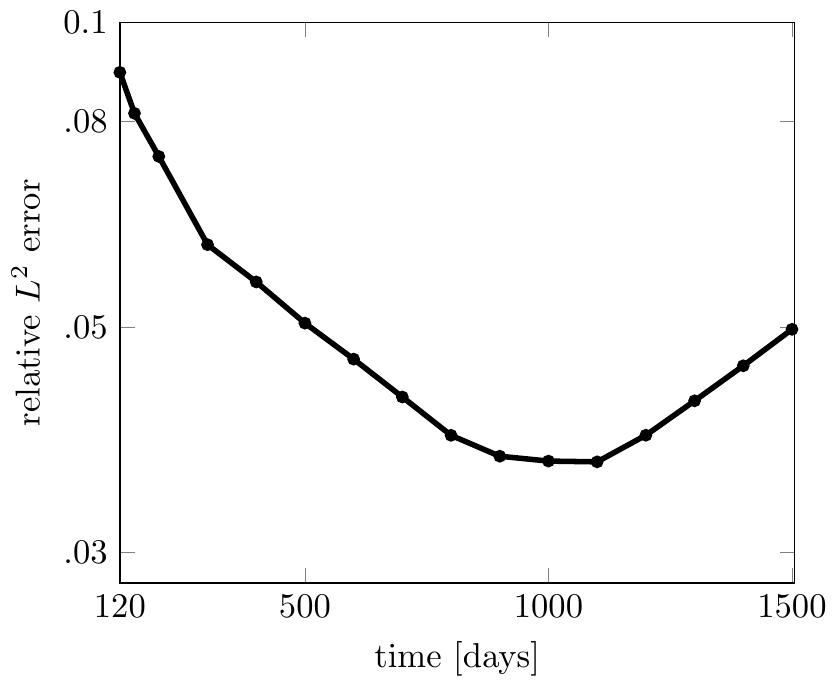}
    \caption{Left: relative $L^2$ error for the log- $\Tarrival$, $\Qmax$, $\Tmaxleak$, as PC order increases;
    middle: time evolution of relative $L^2$ error for the log-$\Qleak$; right: relative $L^2$ error 
     of 4th order PC expansion for $\Qleak$ over time.} 
    \label{fig:L2error}    
\end{figure}

The accuracy of our spectral representations, so far, have been
examined by looking at the convergence of the pdfs, indicating
convergence in distribution.
To further examine the accuracy of the spectral representation of the QoIs, we
approximate relative $L^2$ errors defined in~\eqref{sec:L2_error}.  To avoid
computing these errors over the same $625$ quadrature nodes used to construct
the PC representations, we approximated the relative $L^2$ errors by quadrature, 
in the lower resolution grid with four points in each stochastic dimension. Since the
quadrature grids are not nested, this approach provides a reliable cross
validation of the computed PC representations. 
Figure~\ref{fig:L2error}~(left) shows the relative $L^2$ error for the scalar
quantities of interest, $\Tarrival$, $\Qmax$, $\Tmaxleak$, versus the PC order,
and hence the number of quadrature nodes required for computing the PC
expansion, increases.  Figure \ref{fig:L2error}~(middle) shows the time evolution of
the relative $L^2$ errors for $t > 120$, at which time over $98\%$ of realizations of
the model output indicate arrival of $\CO$ plume to the leaky well.  

From Figure~\ref{fig:L2error}~(left), we note that with a fourth-order PC
expansion, the errors are at around $1\%$ for log of $\Tarrival$ and $\Qmax$,
and around $2\%$ for log $\Tmaxleak$; the errors are acceptable even for a
third-order expansion.  Moreover, from Figure~\ref{fig:L2error}~(middle), we
note that after the initial transient regime, the error for the fourth-order PC
expansion for log-$\Qleak$ is about $1\%$. As before, we also note that
a third-order PC expansion provides a good balance between accuracy and
computational cost.  

The relative $L^2$ errors reported correspond to the PC expansion for the log of the
quantities of interest.%
We also examined the $L^2$ error of the computed quantities of
interest: with a fourth-order PC expansion, the estimated
relative $L^2$ errors for $\Tarrival$, $\Qmax$, and $\Tmaxleak$ were about
$4\%$, $2\%$, and $15\%$, respectively. Moreover, the relative $L^2$ error for
the $\CO$ leakage over time was no more than around $8.5\%$ for $t > 120$. 

The idea of projecting the log of a quantity in a PC basis and approximating it
by evaluating the exponential of the PC expansion was found to be a useful tool
in simulating the distribution of the quantities of interest in the present
study---it was observed to improve convergence in distribution as well as preserving positivity 
of quantities of interest. However, we found that projecting the $\CO$ leakage directly into PC
basis provides acceptable accuracy also (see Figure~\ref{fig:L2error}~(right)),
and is convenient to use for global sensitivity analysis, presented later in
this section. 
 
We see that in this problem, the error in PC representation stabilizes over time, and a
fixed (low) order PC expansion is suitable over the simulation time as the system
tends to an equilibrium. While this phenomenon holds in many applications, we
point out that there are important situations where a straightforward
application of PC methodology is not suitable and one needs to resort to
techniques such as preconditioning~\cite{alen:2012}, asynchronuous
integration~\cite{maitre:2010}, or techniques such as ones proposed
in~\cite{Gerritsma10}, when constructing spectral representations over time.  A
challenge that could occur in some problems (not observed in the present study)
is a form of parametric stiffness that entails excitation of higher-order PC
coefficients over time, entailing the need for increasing the PC order as time
increases; see e.g., the discussion in~\cite{alen:2012,AlexanderianRizziRathinamEtAl14}.

\begin{figure}[ht]\centering
\begin{tabular}{cc}
    \includegraphics[width=0.33\textwidth]{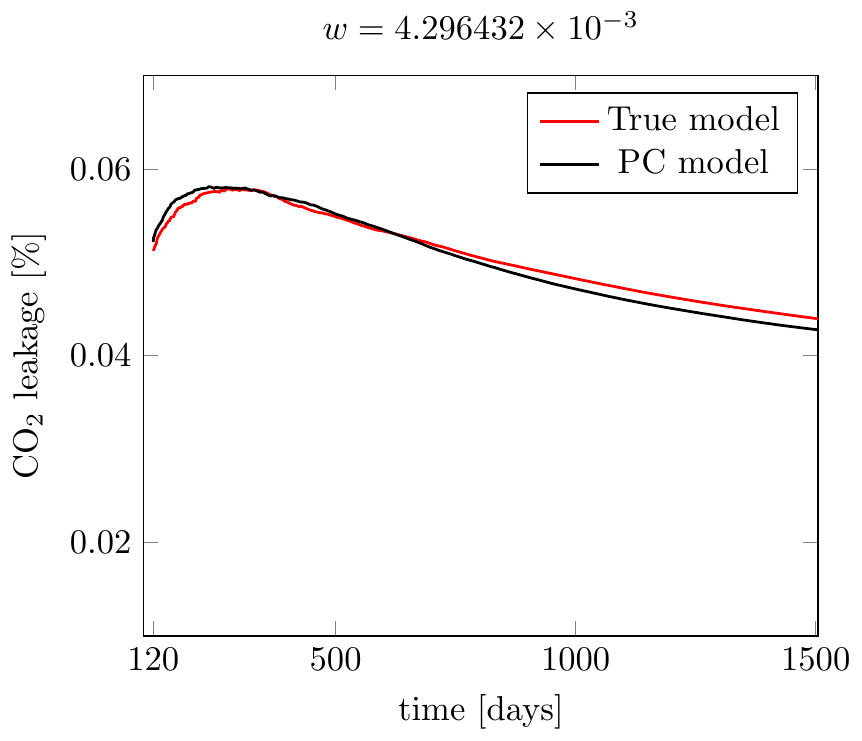}&
    \includegraphics[width=0.33\textwidth]{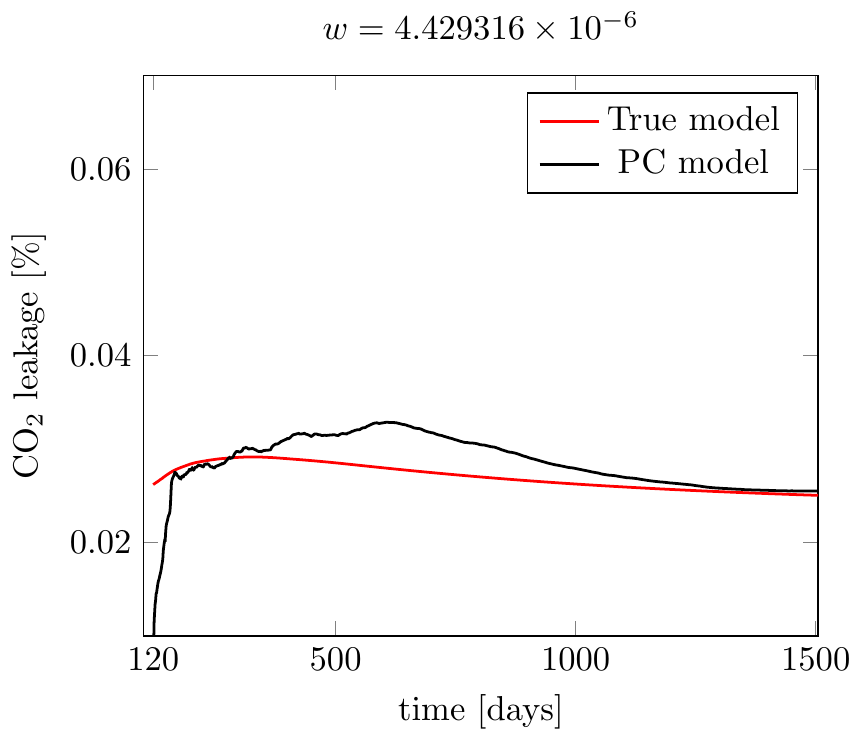}\\
    (a) & (b) \\
    \includegraphics[width=0.33\textwidth]{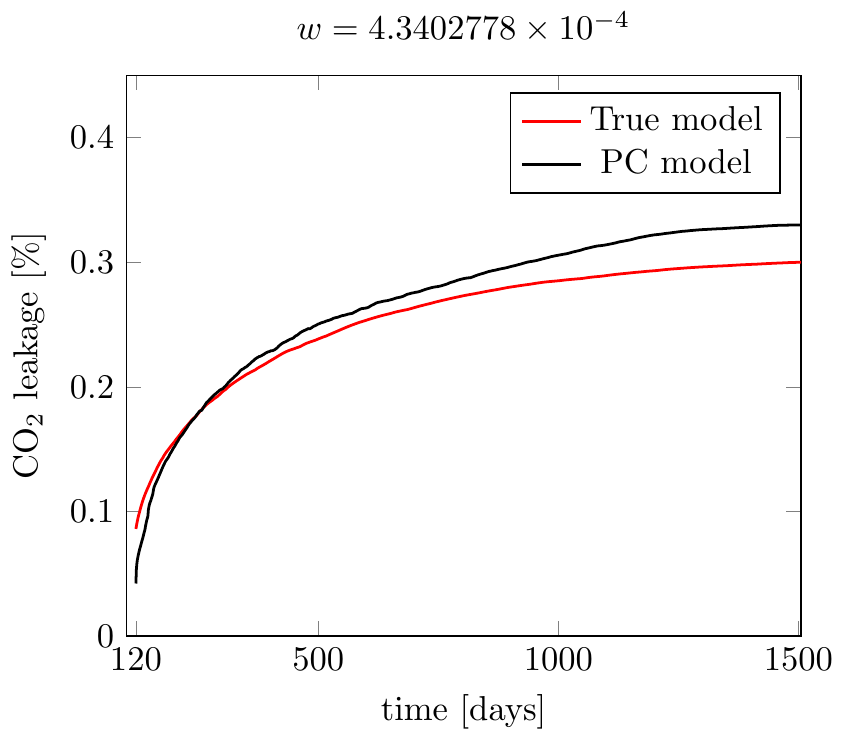}&
    \includegraphics[width=0.33\textwidth]{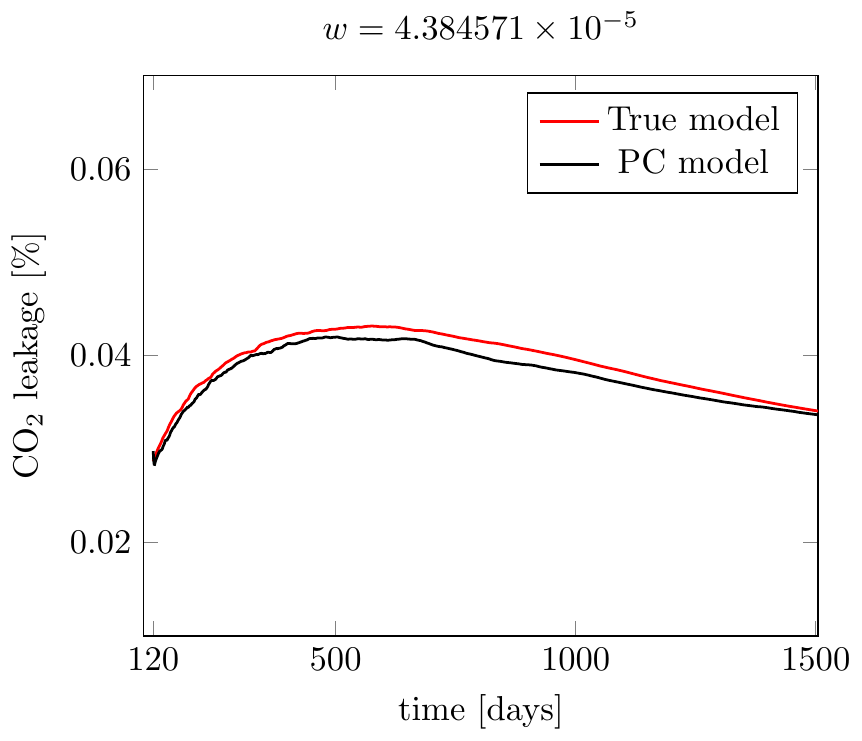}\\
    (c) & (d)
\end{tabular}
\caption{Four realizations of the leakage computed using the numerical
model (red) and the corresponding realization using the PC model (black).}
\label{fig:rlz}
\end{figure}

We also provide a visual comparison between the ``true'' $\CO$ leakage, computed with
the numerical model, and its approximation given by the PC model in
Figure~\ref{fig:rlz}.  We selected four realizations of the $\CO$ leakage, from
among the $256$ realizations, which we used as validation data in the error
study above. For each realization we evaluate the PC model at the same
point. The plots show a best case (Figure~\ref{fig:rlz}(a)), where the PC model agrees well with the
numerical model, and a worst case (Figure~\ref{fig:rlz}(b)), and two other
realizations (bottom row). We also report the quadrature weight corresponding to each point
in the title of each plot. Note that the worst case scenario corresponds to
the smallest quadrature weight from among the ones reported.  To provide
physical insight, we also report the parameter values corresponding to each
plot in Table~\ref{tbl:rlz}.  Note that the realization in
Figure~\ref{fig:rlz}(c) exhibits a very different physical response compared to
the other cases. This is due to the fact that this realization
corresponds to a very low absolute permeability in the formation compared to leaky well
permeability, which entails large levels of $\CO$ leakage.
\begin{table}[ht]
\caption{Parameter values corresponding to the realizations in 
Figure~\ref{fig:rlz}, rounded to two decimal places. Here $\phi$ is porosity, $K_A$ is absolute permeability, 
and $K_L$ is leaky well permeability.}
\label{tbl:rlz}\centering
\begin{tabular}{l|llll}
Image & $\phi$ [-] & $K_A$ [mD] & $K_L$ [mD] & Injection rate [kg/s] \\ 
\hline
(a) & $1.29\times10^{-1}$ & $230.52$ & $2391.78$ & $7.65$ \\
(b) & $9.40\times10^{-2}$ & $1558.16$ & $2391.78$ & $14.15$ \\ 
(c) & $2.39\times10^{-1}$ & $5.75$ & $1331.34$ & $7.65$\\
(d) & $2.39\times10^{-1}$ & $38.85$ & $429.29$ & $14.15$ \\
\end{tabular}
\end{table}

We emphasize that whereas a single evaluation of the numerical model took 18
hours (using five cores), as mentioned above, evaluating a PC expansion has
trivial computational cost (less than a second). Our results
highlight the efficiency of PC expansions in building reasonably accurate and
cheap-to-evaluate surrogate models.

\subsection{Studying the impact of quadrature}

\begin{figure}[h]\centering
    \begin{tabular}{cc}
    \includegraphics[width=0.35\textwidth]{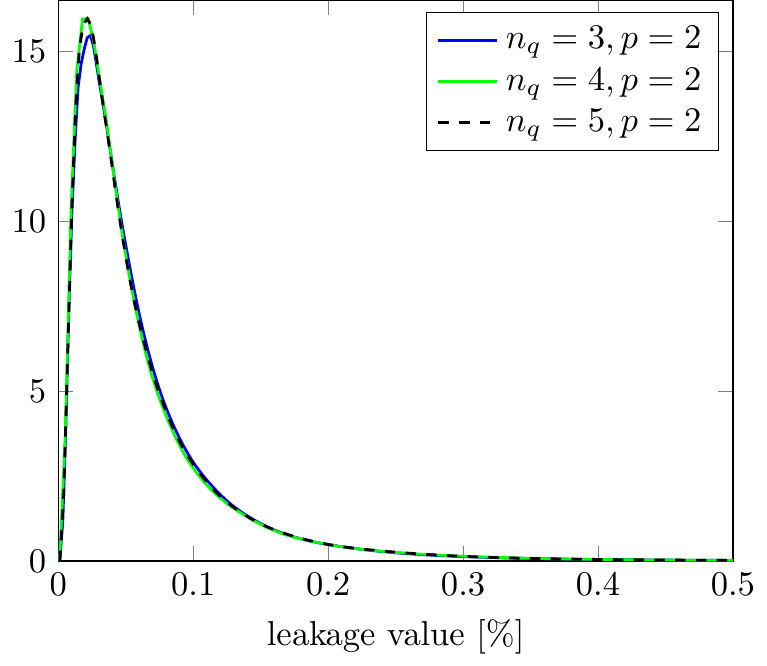} &
    \includegraphics[width=0.35\textwidth]{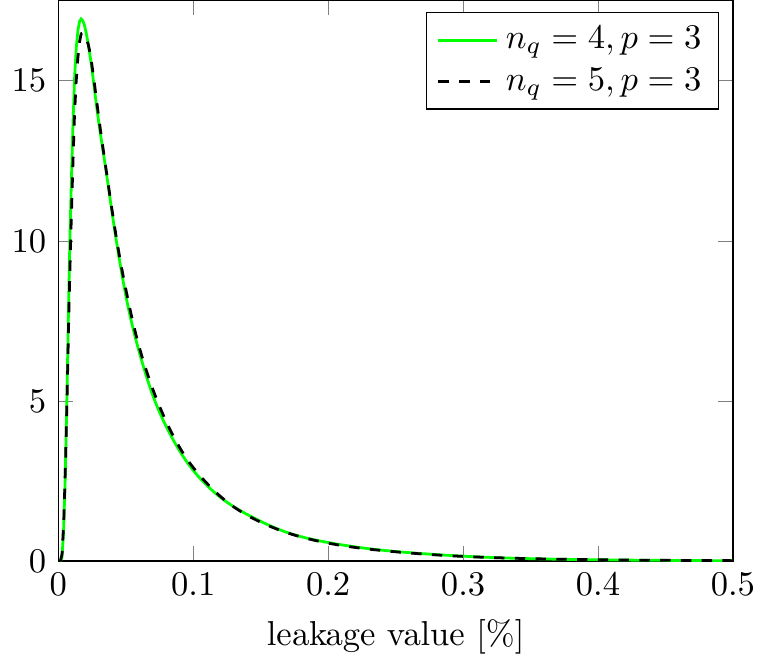}
    \end{tabular}
    \caption{The effect of quadrature on distribution of maximum leakage as approximated by PC 
             expansions of order 2 (left) and order 3 (right). Here $n_q$ denotes the number of 
             Gauss--Hermite quadrature points in each stochastic dimension. 
             Note that in each case, the total number of quadrature nodes is $n_q^4$.}
    \label{fig:quadrature_study}
\end{figure}

Our choice of the resolution of the quadrature rule is mainly guided by the requirement that
the quadrature formula should preserve the discrete orthogonality of the PC basis. 
That is, we require that (within machine precision)
\[
   \ip{ \Psi_i }{\Psi_j} \approx \sum_{k = 1}^{N_q} \Psi_i(\xi_k) \Psi_j(\xi_k)
w_k = \delta_{ij} \ip{ \Psi_i }{\Psi_i} . \] 
This has guided the number of quadrature nodes we have placed in each
stochastic dimension. However, for a given QoI $X$, the computation of PC modes
requires evaluating~\eqref{eq:Integration} whose accuracy will be affected by
regularity of $X$. Hence, to gain confidence in our computations, we need to
examine the effect of the resolution of the quadrature on the PC representation
of the QoIs. In Figure~\ref{fig:quadrature_study}, we study this by looking at
the distribution of the maximum leakage, which is a key QoI, when sampling its
PC expansion of order two and three, computed using quadrature formulas with 
increasing resolutions.  These results indicate that the choice of
the quadrature is appropriate to compute the PC expansions, for maximum
leakage. Similar behavior was observed with other quantities of interest. 

\subsection{Estimating probability of excess $\CO$ leakage}\label{subsec:exceeds_leakage}
\begin{figure}[ht]
  \centering
    \includegraphics[width=0.425\textwidth]{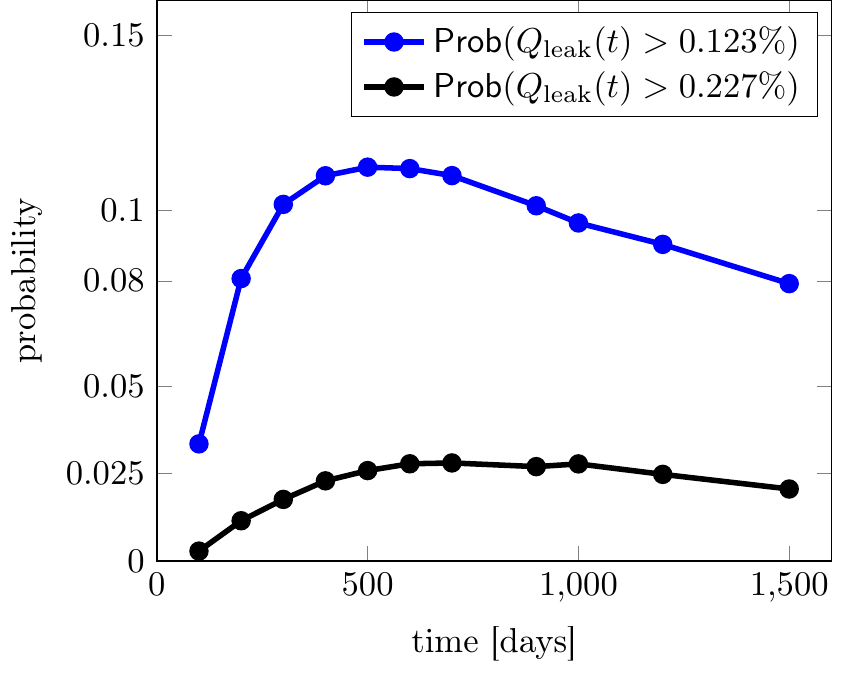}
    \includegraphics[width=0.45\textwidth]{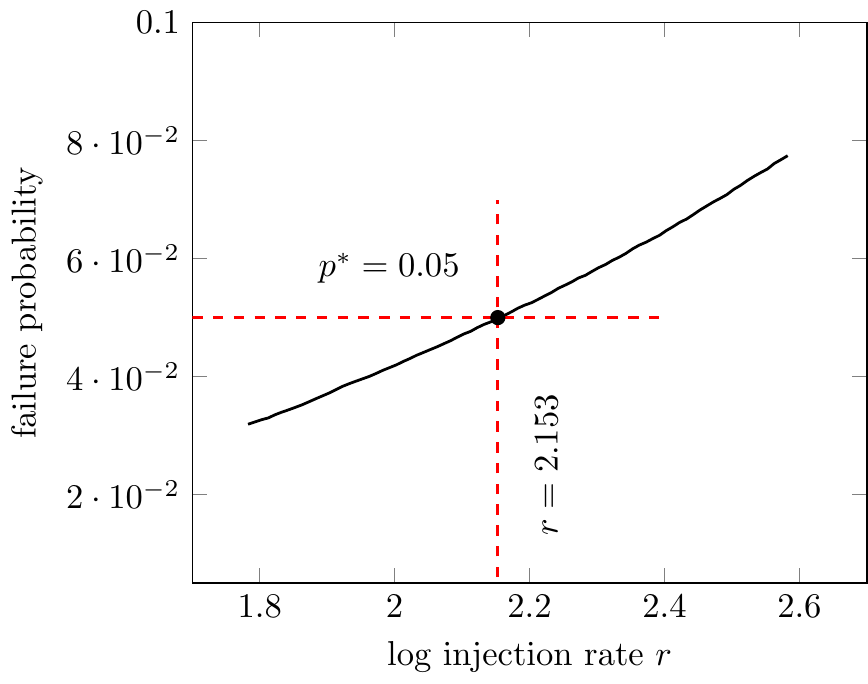}
\caption{
Left: Probability of excess leakage, as defined by leakage
exceeding $0.123\%$ (blue) and $0.227\%$ (black), over time. 
Right:Curve showing $\Pi_\text{fail}(r)$ as a function of $r$.  We
notice that $r \approx 2.153$ gives the largest log-injection rate (which translates 
to an injection rate of approximately $8.61$ kg/s) that
  results in probability of failure being less than $5\%$.}
\label{fig:risk_analysis}
\end{figure}

An important consideration in modeling $\CO$ leakage in reservoirs is
understanding the likelihood of excess $\CO$ leakage.  In particular,
we consider the probability, \newcommand{\prob}{\mathrm{prob}}
\[
P_\text{excess leakage}(t) := \prob( \Qleak(t, \xxi) > L_\text{max}), 
\]
over time.  Notice that computing such a probability is in general a
computationally expensive task. However, using the cheap-to-evaluate PC
representation of $\Qleak$ enables estimation of such probabilities at
negligible computational cost.  Below we use $L_\text{max} = 0.123\%$ and
$L_\text{max} = 0.227\%$, which  correspond, respectively, to the maximum leakage
ratio obtained in the benchmark study of $\CO$ leakage through an abandoned
well \cite{Bench1} using a simplifying assumption to reduce the complexity of
the equations to that obtained by the more physically detailed equations used also in our
study. Figure~\ref{fig:risk_analysis}~(left) depicts the time-dependent behavior of
$P_\text{excess leakage}(t)$.  These results indicate that given our assumed
statistical model for the uncertain parameters, the probability of $\Qleak$
exceeding $0.227\%$ remains below $5\%$, but probability of $\Qleak$ exceeding
$0.123\%$ reaches values of more than $10\%$.

\subsection{Choosing optimum design to reduce risk of failure}\label{subsec:caprock_failure}
Here we consider the caprock
pressure at a point near the leaky well, and denote this quantity by
$p_\text{caprock}(t, \xi)$. Let us consider the quantity,
\[
   p^*_\text{caprock}(\xi) = p_\text{caprock}(t=1500, \xi).
\]

We define the failure
probability as that of $p^*_\text{caprock}$ exceeding a critical
caprock pressure equal to $330$ bar.
An optimal injection rate is the largest injection rate for which the failure
probability remains below $5$ percent. To define this quantity mathematically,
we denote
\[
\Pi_\text{fail}(Q_{\CO}) := \text{Prob}(p^*_\text{caprock}(\xi_1, \xi_2, \xi_3;  \xi_4(Q_{\CO})) > 330),
\quad \text{with }
\xi_4(Q_{\CO}) = (Q_{\CO} - {\bar{Q}_{\CO}})/\sigma_{Q_{\CO}},
\]
where $\bar{Q}_{\CO}$ and $\sigma_{Q_{\CO}}$ are the mean and standard deviation of $Q_{\CO}$ 
chosen according to Table 2. 
Note that to compute $\Pi_\text{fail}(Q_{\CO})$ at a given rate $Q_{\CO}$, we use the PC surrogate for $p^*_\text{caprock}$, 
fix $\xi_4$ at $\xi_4(Q_{\CO})$ and, considering $p^*_\text{caprock}$ as a function of $\xi_1$, $\xi_2$, $\xi_3$, 
compute the probability $p^*_\text{caprock} > 330$ bar using Monte Carlo Sampling, which can be done very 
efficiently %
using the PC surrogate.
Then, we define the optimum injection rate $Q_{\CO}^*$ according to 
\[
    Q_{\CO}^* = \max_{Q_{\CO}} \{Q_{\CO} : \Pi_\text{fail}(Q_{\CO}) < 0.05\}.
\]
Figure \ref{fig:risk_analysis}~(right) illustrates the choice of the injection
rate based on a critical caprock pressure equal to $330$ bar after
$1500$ days. The results in Figure \ref{fig:risk_analysis}~(right) indicate
that the maximum injection rate where the caprock pressure does not
exceed the limit of $330$ bar is approximately $8.61$ kg/s.

\subsection{Global sensitivity analysis}\label{sec:sensitivity_analysis}
In this section, we analyze the
importance of each of the uncertain parameters to the uncertainties in
the $\CO$ leakage. To this end, we perform a variance-based
sensitivity analysis, where we find how much each uncertain input
parameter contributes to the total variance in $\CO$ leakage. 

\begin{figure}[ht]
\centering
\includegraphics[width=0.4\textwidth]{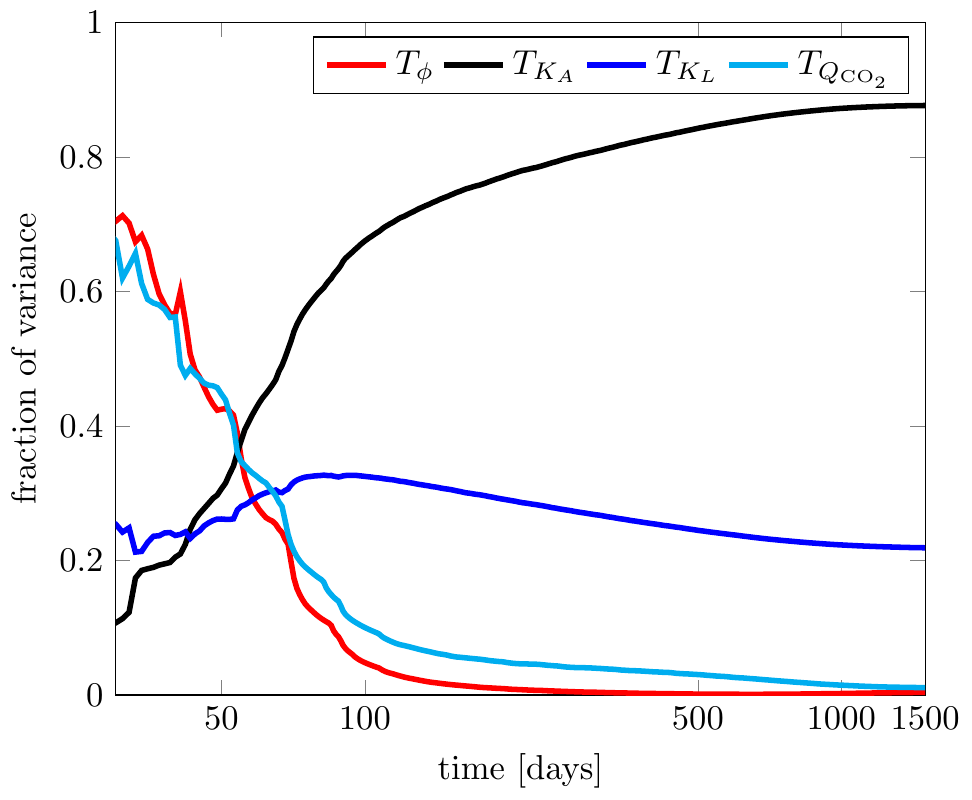}
\includegraphics[width=0.4\textwidth]{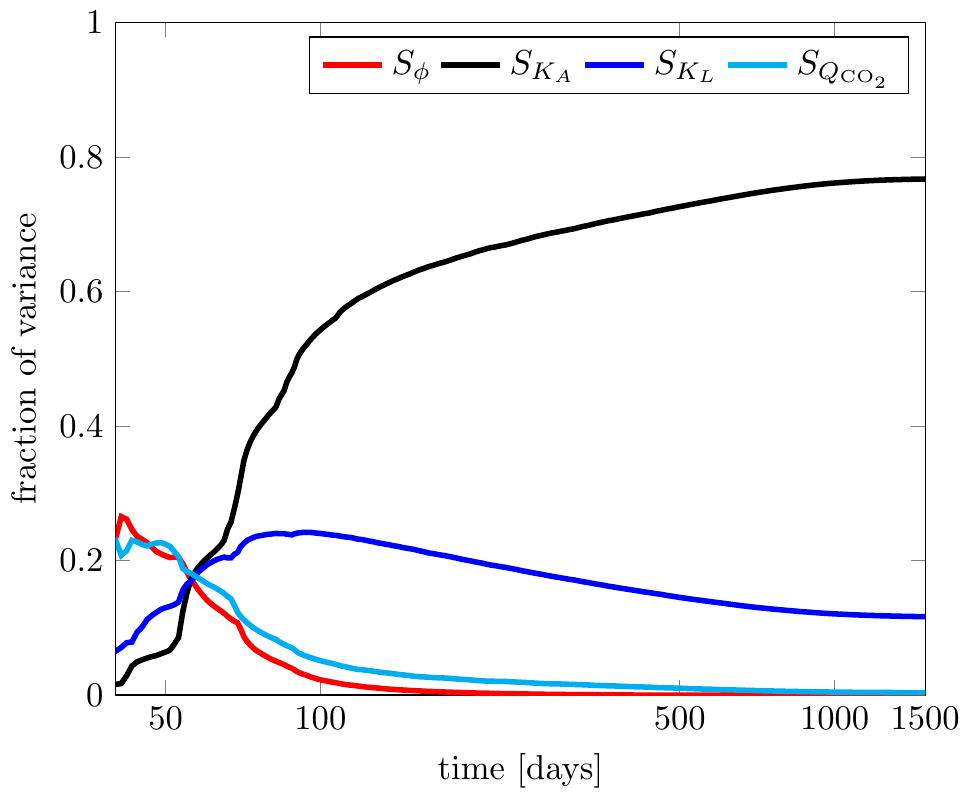}
\caption{
Variance based sensitivity analysis for $\CO$ leakage, $\Qleak$.
Left: total sensitivity indices over time. Right: first-order sensitivity indices over time.}
\label{fig:total_1st_order_SA}
\end{figure}

Figure \ref{fig:total_1st_order_SA}~(left) depicts the time-dependent behavior
of total sensitivity indices for $\CO$ leakage. For clarity, we denote the
total sensitivity indices for the random inputs by $T_\phi$, $T_{K_A}$,
$T_{K_L}$, and $T_{Q_{\CO}}$, corresponding to total sensitivity index for
porosity, reservoir permeability, leaky well permeability and $\CO$ injection
rate.  The results in Figure~\ref{fig:total_1st_order_SA}~(left) indicate that
at early time, the porosity and the $\CO$ injection rate have a significant
influence on the total variance. However, as the flow reaches the leaky well,
the variance of $\CO$ becomes dominated by the uncertainties in $K_A$ and
$K_L$.  

\begin{figure}[ht]
\centering
\includegraphics[width=0.6\textwidth]{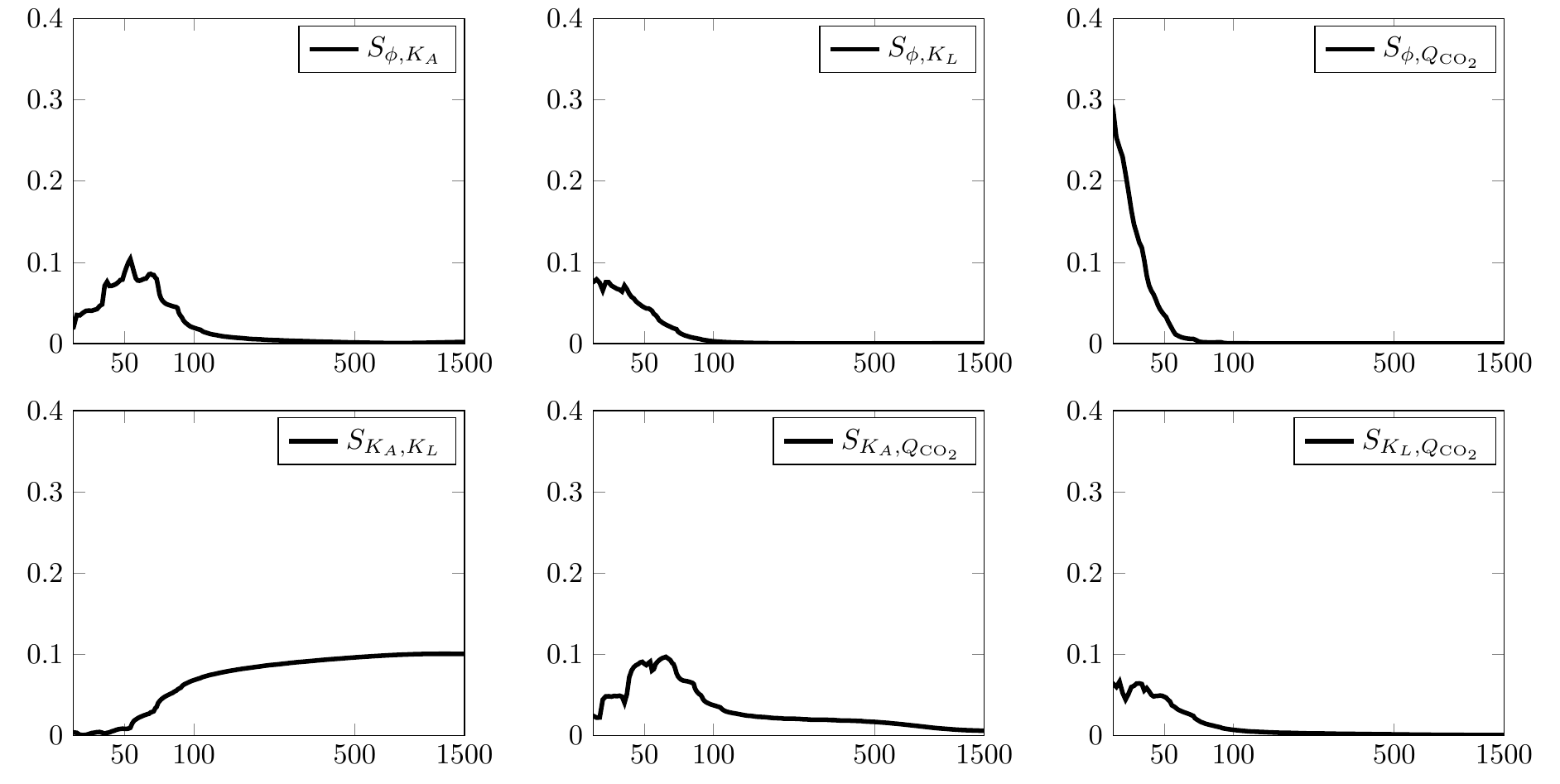}
\includegraphics[width=0.32\textwidth]{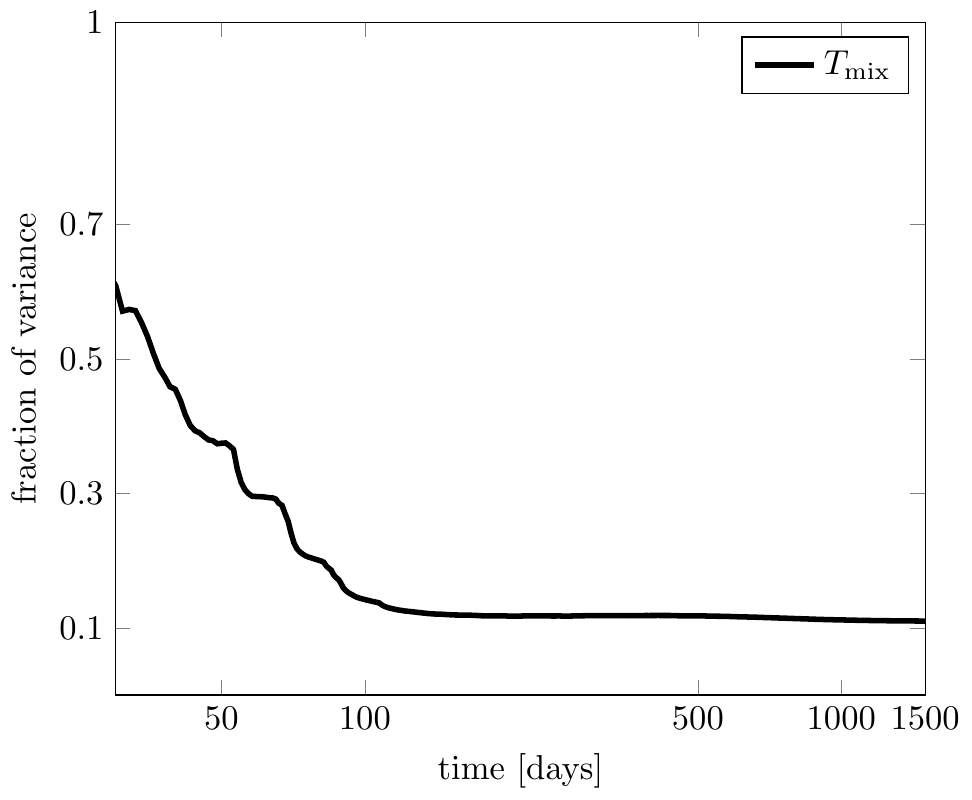}
\caption{
Variance based sensitivity analysis for $\CO$ leakage, $\Qleak$.
Left: second-order sensitivity indices over time. 
Right: the mixed sensitivity index $T_\text{mix}$ over time. }
\label{fig:mixed_effects}
\end{figure}

We also report the first-order indices $S_\phi$, $S_{K_A}$, $S_{K_L}$, and
$S_{Q_{\CO}}$ in Figure~\ref{fig:total_1st_order_SA}~(right).  While the first-order
indices show a similar trend as the total sensitivity indices, comparing
their values with total indices, especially at early times, suggests that
interactions between parameters have a significant contribution to variability
in $\CO$ leakage.  To further understand this, we compute the second-order
indices, which quantify the contribution of the interaction between random
parameters to the total variance; the results are reported in
Figure~\ref{fig:mixed_effects}~(left).  We note that at early times the interactions
between the uncertain parameters have a noticeable contribution to the total
variance, but as the time passes the second-order interactions mostly vanish
(except the interaction between $K_A$ and $K_L$) and the first-order indices
are almost equal to the total indices reported in
Figure~\ref{fig:total_1st_order_SA}~(left). That is, as the $\CO$ plume reaches
the leaky well the response of the system becomes nearly additive in the
uncertain parameters.

To quantify the total contribution of the interactions between uncertain 
parameters to the variance of model output ($\CO$ leakage in the present case), 
we consider the following \emph{mixed sensitivity index}, 
\[
   T_\text{mix} \defeq \frac{\text{variance due to interactions between parameters}}{\text{total variance}}.
\]
This mixed index can be defined in terms of conditional expectations (cf. Section~\ref{sec:SA}), 
and is general difficult to approximate. However, using the PC representation of the model response
($\CO$ leakage here), we can easily derive the following expression for $T_\text{mix}$. %
Using the multi-index construction of the multivariate PC basis in~\eqref{equ:multivariatePC}, 
we can define,
\[
    T_\text{mix} \approx \frac{\sum_{k \in \mathcal{K}} c_k^2 \norm{\Psi_k}^2}
    {\sum_{k = 1}^P c_k^2 \norm{\Psi_k}^2},
\]
where the index set $\mathcal{K}$ is defined by 
\[
   \mathcal{K} = \{ k \in \{1, \ldots, P\} : \| \vec{\alpha}_k \|_0 > 1 \}.
\]
Note that here we have used the multi-index notation, used in construction of the 
PC basis, and denoted by $ \| \cdot \|_0$ the $\ell_0$-``norm''. That is, 
for a vector $\vec{x}$, $ \| \vec{x} \|_0$ is the number of nonzero elements of 
$\vec{x}$.
The index $T_\text{mix}$ quantifies the contribution to the variance due to all
interactions (second- and higher-order) between the uncertain inputs. Using the
fourth-order PC expansion we have computed for $\CO$ leakage, we approximate the
mixed index for this QoI; see Figure~\ref{fig:mixed_effects}~(right). The results reported in
Figure~\ref{fig:mixed_effects}~(right) show that, as also seen from the study of first- and
second-order indices, there is significant contributions to model variability
coming from interactions between the parameters at early times.  These
mixed-effect interactions level off at around $10\%$ as the $\CO$ plume
reaches the leaky well.

\begin{figure}[ht]
\centering
\includegraphics[width=0.5\textwidth]{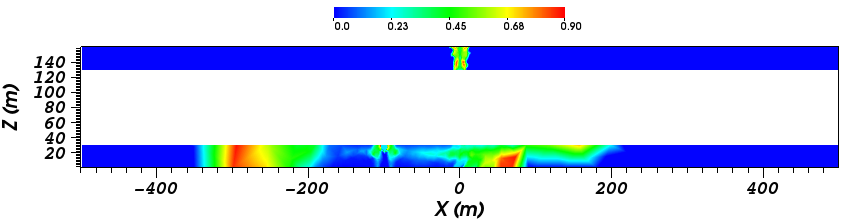}
\includegraphics[width=0.5\textwidth]{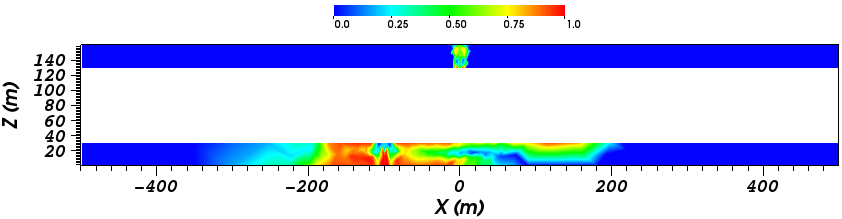}
    \caption{Space propagation of sensitivities after 70
      days. Here we consider the sensitivity of saturation to porosity $\phi$ (top) and reservoir permeability (bottom).} \label{fig:3Dsens}
\end{figure}

Finally, we compute the sensitivity indices over
the three-dimensional computational domain. In particular, we consider the
sensitivity of $\CO$ saturation to the uncertain inputs.  Figure
\ref{fig:3Dsens} shows the spatial distribution of the total sensitivity indices,
$T_\phi$ and $T_{K_A}$ for $\CO$ saturation at $t = 70$ days.
A vertical slice through the middle of the domain indicates that the 
regions where porosity has a significant contribution to variance travel with
the fronts of the $\CO$ plume, whereas the reservoir permeability maintains a
nearly constant dominant effect on variance within the regions with high $\CO$
saturation.

%
%
\section{Conclusions and Summary}\label{sec:conclusion}
A non-intrusive spectral projection approach was implemented to propagate and
quantify parametric uncertainties for $\CO$ storage in geological formations
using a common 3D leakage benchmark problem of injected $\CO$ into overlying
formations through a leaky well.  A non-isothermal two-phase two-component flow
system with equilibrium phase exchange is used.  Moreover, we use nonlinear
functions for the capillary pressure and the relative permeability for each
phase. 

In our numerical results, we find that the use of a nonlinear relative
permeability-saturation relation in our mathematical model leads to an overall
reduced mobility of the flow.  This behavior is seen in the statistical
distribution of the $\CO$ arrival time to the leaky well.  In particular,
tracking the time evolution of the distribution of the $\CO$ leakage, we see
that the leakage rate starts decreasing after the peak of the arrival of the
$\CO$ flux at the leaky well.  This is in contrast with the 
cases where one uses simplified assumptions such as linear
relative permeabilities, which decreases the influence of the viscous forces
in the system and results in early arrival times of the $\CO$ plume at the
leaky well~\cite{Bench1}.

We find, given our assumed statistical distributions for the random inputs,
that the risk of $\CO$ leakage in excess of $0.123\%$ could exceed $10\%$
within the first two years of the simulation. However, this risk falls well
below $5\%$, when we consider a leakage threshold of $0.227\%$.  In our
computation of optimum injection rate, we find that, given our assumed
statistical distributions for the random inputs, an injection rate of $8.61$
kg/s still ensures low risk of failure (defined as excess pressure buildup at
the leaky well).

In our sensitivity analysis, we find that 
the balance of sensitivities changes as a function of time, where the $\CO$
injection rate and porosity exhibit significant contribution to variance at
early times, and become less important as the $\CO$ plume reaches the leaky
well.  On the other hand, the reservoir permeability and leaky 
well permeability become dominant contributors to the variance of $\CO$ leakage at
later times.  We also find that at early times the interactions
among the different uncertain parameters has significant contribution to
variance, but as the $\CO$ plume reaches the leaky well, the bulk effect of
interactions between the parameters to the variance is due to the permeability
of both the reservoir and the leaky well.
The study of sensitivity of saturation to uncertain inputs in the
three-dimensional domain reveals that the regions where porosity has a
significant contribution to the variance travel with the fronts of the $\CO$
plume; however, the reservoir permeability maintains a nearly constant dominant
effect on variance within the regions with high $\CO$ saturation.

\section*{Acknowledgement}
Research reported in this publication was supported by the King Abdullah
University of Science and Technology (KAUST) under the Academic Excellency
Alliance (AEA) UT Austin-KAUST project "Uncertainty quantification for
predictive modeling of the dissolution of porous and fractured media".
Computational resources for the simulations presented in this publication have
been made available by KAUST Research Computing and KAUST SuperComputing Lab.
Bilal Saad is grateful for the support by the Saudi Arabia Basic Industries
Corporation (SABIC). Bilal Saad, Serge Prudhomme, and Omar Knio are also
participants of the KAUST SRI Center for Uncertainty Quantification in
Computational Science and Engineering.

\bibliographystyle{elsarticle-num}

\section*{\refname}

\bibliography{biblio,uqrefs}

\end{document}